\documentclass[10pt]{article}

\usepackage{amsmath}
\usepackage{amssymb}

\usepackage[pdftex,breaklinks,colorlinks]{hyperref}
\hypersetup{  
  pdftitle =
    {An Exact Expression to Calculate the Derivatives of Position-Dependent Observables in Molecular Simulations with Flexible Constraints}
}
\hypersetup{  
    citecolor = blue,
    urlcolor = blue
}

\usepackage{graphicx}

\usepackage{cite}

\usepackage[usenames]{color} 

\usepackage[labelfont=bf,labelsep=period,justification=raggedright]{caption}

\makeatletter
\renewcommand{\@biblabel}[1]{\quad#1.}
\makeatother

\date{}

\begin{document}

\begin{flushleft}
{\Large
\textbf{An Exact Expression to Calculate the Derivatives of Position-Dependent Observables in Molecular Simulations with Flexible Constraints}
}
\\
Pablo Echenique$^{1,2,3,4,\ast}$, 
Claudio N. Cavasotto$^{5,6}$, 
Monica De Marco$^{3}$,
Pablo Garc\'{\i}a-Risue\~no$^{1,3,2}$, 
J.~L.~Alonso$^{3,2,4}$
\\
\bf{1} Instituto de Qu\'{\i}mica F\'{\i}sica ``Rocasolano'', CSIC, Madrid, Spain
\\
\bf{2} Instituto de Biocomputaci\'on y F{\'{\i}}sica de Sistemas Complejos (BIFI), Universidad de Zaragoza, Zaragoza, Spain
\\
\bf{3} Departamento de F{\'{\i}}sica Te\'orica, Universidad de Zaragoza, Zaragoza, Spain
\\
\bf{4} Unidad Asociada IQFR-BIFI
\\
\bf{5} School of Biomedical Informatics, University of Texas Health Science Center at Houston, Houston, USA
\\
$\ast$ E-mail: echenique.p@gmail.com
\end{flushleft}

\section*{Abstract}

In this work. we introduce an algorithm to compute the derivatives of physical
observables along the constrained subspace when flexible constraints are
imposed on the system (i.e., constraints in which the constrained coordinates
are fixed to configuration-dependent values). The presented scheme is exact,
it does not contain any tunable parameter, and it only requires the
calculation and inversion of a sub-block of the Hessian matrix of second
derivatives of the function through which the constraints are defined. We also
present a practical application to the case in which the sought observables
are the Euclidean coordinates of complex molecular systems, and the function
whose minimization defines the flexible constraints is the potential energy.
Finally, and in order to validate the method, which, as far as we are aware,
is the first of its kind in the literature, we compare it to the natural and
straightforward finite-differences approach in a toy system and in three
molecules of biological relevance: methanol, N-methyl-acetamide and a
tri-glycine peptide.

\section*{Author Summary}

\section*{Introduction}
\label{sec:introduction}

In the theoretical and computational modeling of physical systems, including
but not limited to condensed-matter materials \cite{Rap2004Book}, fluids
\cite{All2005Book}, and biological molecules \cite{Fre2002Book}, it is very
common to appeal to the concept of \emph{constraints}. When a given quantity
related to the system under study is constrained, it is not allowed to depend
explicitly on time (or on any other parameter that describes the evolution of
the system in the problem at hand). Instead, a constrained quantity is either
set to a constant value (\emph{hard} or \emph{rigid} constraints) or to a
function of the rest of degrees of freedom (\emph{flexible}, \emph{elastic} or
\emph{soft} constraints); in such a way that, if it depends on time, it does
so through the latter and not in an explicit manner.

The imposition of constraints is useful in a wide variety of contexts in the
fields of computational physics and chemistry: For example, we can use
constraints to maintain an exact symmetry of the equations of motion; like in
Car-Parrinello molecular dynamics (MD) \cite{Car1985PRL}, where the
time-dependent Kohn-Sham orbitals need to be orthonormal along the time
evolution of the quantum-classical system, a requirement that can be fulfilled
by imposing constraints over their scalar product \cite{Hut2005CPC}. In a
different context, we can use constraints, as in the Blue Moon Ensemble
technique \cite{Car1989CPL}, to fix some macroscopic, representative degrees
of freedom of molecular systems (normally called \emph{reaction coordinates}),
in order to be able to compute free energy profiles along them that would take
an unfeasibly long time if we used an unconstrained simulation. Probably the
most common application of the idea of constraints, and the one that will be
mainly discussed in this work, appears when we fix the fastest, hardest
degrees of freedom of molecular systems, such as bond lengths or bond angles,
in order to allow for larger time-steps in MD simulations
\cite{Sch1997ARBPBMS,Ryc1977JCoP}.

In any of these cases (assuming that the dimensions of the spaces involved are
all finite) the imposition of constraints can be described in the following
way: If the state of the system is parameterized by a given set of coordinates
$q := (q^\mu)_{\mu=1}^N$, spanning the \emph{whole space}, $\mathcal{W}$, and
the associated momenta $p := (p_\mu)_{\mu=1}^N$, a given \emph{constrained
subspace}, $\mathcal{K}$, of dimension $K<N$, can be defined by giving a set
of $L:=N-K$ independent relations among the coordinates (In this work, we will
only deal with holonomic, scleronomous constraints, i.e., those that are
independent both of the momenta and (explicitly) of time.):
\begin{equation}
\label{eq:constraints1}
h^I(q) = 0  \ , \quad I=K+1,\ldots,N \ .
\end{equation}

The condition of these constraints being independent amounts to asking the set 
of $L$ vectors of $N$ components
\begin{equation}
\label{eq:gradientsh}
\partial h^I(q) := \left( \frac{\partial h^I}{\partial q^\mu}(q)
  \right)_{\mu=1}^N , \quad I=K+1,\ldots,N \ ,
\end{equation}
to be linearly independent at the relevant points $q$
satisfying~(\ref{eq:constraints1}), and it means that $\mathcal{K}$ is a
manifold of constant dimension in these points, which are called
\emph{regular}. Moreover, this independence condition allows, in the vicinity
of each point $q$ and by virtue of the Implicit Function Theorem
\cite{Dub1992Book,Wei2009Web1}, to (formally) solve~(\ref{eq:constraints1})
for $L$ of the coordinates, which we arbitrarily place at the end of $q$,
splitting the original set as $q=(u,d)$, with $u := (u^r)_{r=1}^K$ and $d :=
(d^I)_{I=K+1}^N$. Then, in the vicinity of each point $q$
satisfying~(\ref{eq:constraints1}), we can express the relations defining the
constrained subspace, $\mathcal{K}$,
\emph{parametrically} by
\begin{equation}
\label{eq:constraints2}
d^I = f^I(u) \ , \quad I=K+1,\ldots,N \ .
\end{equation}
where the functions $f(u) := \big(f^I(u)\big)_{I=K+1}^N$ are the ones whose
existence the Implicit Function Theorem guarantees. The coordinates $u$ are
thus termed \emph{unconstrained} and they parameterize $\mathcal{K}$, whereas
the coordinates $d$ are called \emph{constrained} and their value is
determined at each point of $\mathcal{K}$ according
to~(\ref{eq:constraints2}). In general, the functions $f^I$ will depend on
$u$, and the constraints will be said to be \emph{flexible}
\cite{Ech2011Sub2}. In the particular case in which all the functions $f^I$
are constant along $\mathcal{K}$, the constraints are called \emph{hard}, and
all the calculations are considerably simplified. In this work, we tackle the
general, more involved, flexible case.

Of course, even if $\mathcal{K}$ is regular in all of its points, the
particular coordinates $d^I$ that can be solved need not be the same along the
whole space. One of the simplest examples of this being the circle in
$\mathbb{R}^2$, which is given by $f(x,y):=x^2+y^2-R^2=0$, an implicit
expression whose gradient is non-zero for all $(x,y) \in \mathcal{K}$.
However, if we try to solve, say, for $y$ in the whole space $\mathcal{K}$, we
will run into trouble at $y=0$; if we try to solve for $x$, we will find it to
be impossible at $x=0$. I.e., the Implicit Function Theorem does guarantee
that we can solve for \emph{some} of the original coordinates at each regular
point of $\mathcal{K}$, but sometimes the solved coordinate has to be $x$ and
sometimes it has to be $y$. Nevertheless, we will assume this to be the case
throughout this work, as is normally done in the literature
\cite{Ech2006JCC2,Chr2005JCP,Hes2002JCP,Zho2000JCP}, and thus we will consider
that $\mathcal{K}$ is parameterized by the \emph{same} subset of coordinates
$u$ for all of its points.

It is also worth mentioning at this point that, not only from the physical
point of view all the constraints dealt with in this work are just holonomic
constraints, but also the wording used to refer to the two \emph{flexible} and
\emph{hard} sub-types is multiple in the literature. The first sub-type is
called \emph{flexible} in
refs.~\cite{Chr2007CPC,Chr2005JCP,Hes2002JCP,Zho2000JCP}, \emph{elastic}
in~\cite{Cot2004BITNM}, and \emph{soft} in~\cite{Zho2000JCP}; whereas the
second sub-type is called \emph{hard} in refs.~\cite{Chr2007CPC,Zho2000JCP},
just \emph{constrained} in~\cite{Chr2005JCP}, or \emph{holonomic}
in~\cite{Cot2004BITNM}, \emph{rigid} in~\cite{Hes2002JCP,Zho2000JCP}, and
\emph{fully constrained} in~\cite{Zho2000JCP}. Some of these terms are clearly
misleading (\emph{elastic}, \emph{holonomic} or \emph{fully constrained}),
and, in any case, so many names for such simple concepts is detrimental to
understanding in the field.

The situation is further complicated by the fact that, when studying the
statistical mechanics of constrained systems, one can think about two
different models for calculating the equilibrium probability density, whose
names often collide with the ones used for defining the type of constraints
applied. On the one hand, one can implement the constraints by the use of very
steep potentials around the constrained subspace; a model sometimes called
\emph{flexible} \cite{Hel1979JCP,Pec1980JCP}, sometimes called \emph{stiff}
\cite{Ech2006JCC2,vKa1984AJP}. On the other hand, one can assume the
D'Alembert principle \cite{Gol2002Book} and hypothesize that the forces are
just the ones needed for the system to never leave the constrained subspace
during its dynamical evolution; a model normally called \emph{rigid}
\cite{Ech2006JCC2,Hel1979JCP,Pec1980JCP}. The two statistical mechanics models
have long been recognized to present different equilibrium probability
distributions \cite{Go1976MM,Hel1979JCP,Pec1980JCP,vKa1984AJP}, and this is
the major concern in the literature when discussing them. In
refs.~\cite{Ech2011Sub2,Ech2006JCC2}, the reader can find a very detailed
discussion of this issue, which we only touch here briefly for completeness.

It is worth remarking that the two types of constraints and the two types of
statistical mechanics models can be independently combined; one can have
either the \emph{stiff} or the \emph{rigid} model, with either \emph{flexible}
or \emph{hard} constraints, hence making any interference between the two sets
of words undesirable. The wording chosen is this work is, on the one hand,
fairly common, and on the other hand, non-misleading.

Now, if we take any physical observable $X(q)$, depending only on the
coordinates (not on the momenta), and originally defined on the whole space,
$\mathcal{W}$, its \emph{restriction} to $\mathcal{K}$ is given by
\begin{equation}
\label{eq:restrictionX}
Z(u) := X\big(u,f(u)\big) \ ,
\end{equation}
where the symbol has been deliberately changed in order to indicate that
$Z$ and $X$ are different functions.

The derivatives of this observable along $\mathcal{K}$ are thus
\begin{equation}
\label{eq:derZ1}
\frac{\partial Z}{\partial u^r}(u) = 
  \frac{\partial X}{\partial u^r} \big(u,f(u)\big) + 
  \frac{\partial X}{\partial d^I} \big(u,f(u)\big)
    \frac{\partial f^I}{\partial u^r}(u) \ ,
\end{equation}
where we have assumed the convention that repeated indices (like $I$ above)
indicate a sum over the relevant range, and we have omitted (as we will often
do) the range of variation of the index $r$.

In the case of hard constraints, i.e., when the functions $f^I$ are
all constant numbers $d_0^I$, the above expression reduces to
\begin{equation}
\label{eq:derZnonflex}
\frac{\partial Z}{\partial u^r}(u) = 
  \frac{\partial X}{\partial u^r} (u,d_0) \ ,
\end{equation}
where $X(q)$ must be a known function of $q$ (in order to have a well-defined
problem), and its derivative is typically easy to compute. However, if the
constraints are of the more general, flexible form (the ones tackled in this
work), the calculation of the partial derivatives $(\partial f^I / \partial
u^r)(u)$ cannot be avoided.

If the constraints are assumed to be flexible, it is common in the literature
of molecular modeling to define these functions $f^I(u)$ as the values taken
by the coordinates $d$ if we minimize either the total or the potential energy
with respect to all $d^I$ at fixed $u$
\cite{Ech2006JCC2,Chr2005JCP,Hes2002JCP,Zho2000JCP,Ech2011Sub2}. Since the
energy functions used in molecular simulation are typically rather
complicated, such as the ones in classical force fields, with a large number
of distinct functional terms
\cite{Bro2009JCC,Jor1988JACS,Jor1996JACS,Pon2003APC,AMBER10,Pea1995CoPC}, or
the effective nuclear potential arising from the solution of the electronic
Schr\"odinger equation in the ground-state Born-Oppenheimer approximation
\cite{Ech2006JCC2,Ech2007MP}, the minimization of the energy with respect to
the coordinates $d$ has to be performed numerically. Hence, the functions
$f^I(u)$, which are the output of this process, do not have a compact
analytical expression that can be easily differentiated to include it in
eq.~(\ref{eq:derZ1}) (this is even the case in very simple toy systems; see
the Results and Discussion section).

In this work, we present a parameter-free, exact algorithm (up to machine
precision) to calculate the derivatives $(\partial f^I / \partial u^r)(u)$ in
such a case. Although several methods exist in the literature
\cite{Chr2005JCP,Hes2002JCP,Zho2000JCP} for performing MD simulations with
flexible constraints, nobody has dealt, as far as we are aware, with the
computation of these derivatives. Since the general idea can be applied to any
situation in which (1) we have flexible constraints, (2) that are defined in
terms of the minimization of some quantity with respect to the constrained
coordinates, we first introduce, the essential part of the algorithm based on
these two points. Then, we develop a more sophisticated application of this
idea to the calculation of the derivatives along the constrained subspace of
the Euclidean coordinates of molecular systems; a problem that we faced in our
group when trying to calculate the correcting terms associated to mass-metric
tensor determinants that appear in the equilibrium probability density when
constraints are imposed \cite{Ech2006JCC2,Ech2006JCC3,Ech2011Sub2}. Finally,
we perform a comparison between the results obtained with our exact algorithm
and the calculation of the derivatives by finite differences; this serves the
double purpose of numerically validating the algorithm and showing the
limitations of the latter method, which needs the tuning of a parameter for
each particular problem.

\section*{Methods}

\subsection*{General Algorithm}

As we mentioned in the Introduction, we assume that we are dealing with a
constrained problem in which the functions $f^I(u)$ in
eq.~(\ref{eq:constraints2}) are defined as taking the values of the
constrained coordinates $d^I$ that minimize a given function,
$V(q)=V(u,d)$, for each fixed $u$, i.e.,
\begin{equation}
\label{eq:minimum1}
V\big(u,f(u)\big) \leq V(u,d) \ , \ \forall d 
 \in \mathcal{D}\big(f(u)\big) \ ,
\end{equation}
where $\mathcal{D}\big(f(u)\big)$ is a suitable open set in $\mathbb{R}^L$
containing the point $f(u)$. Depending on the particular application, one can
ask the minimum that defines the functions $f^I(u)$ to be global or just
local. However, in the cases in which $V$ is the total or the potential energy
of a complex molecular system, it may become very difficult to find its global
minimum (due to the shear number of dimensions of the search space), and the
local choice is the only reasonable one \cite{Ech2011Sub2}.

In order to calculate the derivatives along $\mathcal{K}$ of any physical
observable function of the coordinates $Z(u) := X\big(u,f(u)\big)$, like the
one defined in~(\ref{eq:restrictionX}), we can always follow the
finite-differences approach. However, as we discuss in the Results and
Discussion section, finite differences presents intrinsic inaccuracies which
are difficult to overcome, specially as the system grows larger. Let us now
introduce a different way to calculate $(\partial Z / \partial u^r)(u)$ which
does not suffer from this drawback.

The starting point is eq.~(\ref{eq:derZ1}) in the Introduction, which we copy
here for the comfort of the reader:
\begin{equation}
\label{eq:derZ1bis}
\frac{\partial Z}{\partial u^r}(u) = 
  \frac{\partial X}{\partial u^r} \big(u,f(u)\big) + 
  \frac{\partial X}{\partial d^I} \big(u,f(u)\big)
    \frac{\partial f^I}{\partial u^r}(u) \ .
\end{equation}

As we mentioned, the expression of $X(q)$, as well as the functions $f^I(u)$,
must be known if we wish to have a well-defined constrained problem to begin
with. Therefore, the only objects that remain to be computed are the partial
derivatives $(\partial f^I / \partial u^r)(u)$.

If we assume that we have available some method to check that the order of the
stationary point is the appropriate one (i.e., that it is a minimum, and not a
maximum or a saddle point), we can write a set of equations which are
equivalent to eq.~(\ref{eq:minimum1}), and which (implicitly) define the
functions $f^I(u)$:
\begin{equation}
\label{eq:minimum2}
\frac{\partial V}{\partial d^I}\big(u,f(u)\big) = 0 \ , \qquad
  I=K+1,\ldots,N  \ .
\end{equation}

Now, we can take the derivative of this expression with respect to a given
unconstrained coordinate $u^r$:
\begin{equation}
\label{eq:derderV}
\underbrace{\frac{\partial^2 V}{\partial u^r \partial d^I}
\big(u,f(u)\big)}_{\displaystyle H_{rI}(u)} +
\underbrace{\frac{\partial^2 V}{\partial d^J \partial d^I}
\big(u,f(u)\big)}_{\displaystyle H_{JI}(u)}
\underbrace{\frac{\partial f^J}{\partial u^r}(u)}_{
\displaystyle F^J_r(u)}
= 0 \ ,
\end{equation}
where $H_{\mu\nu}(u)$, with $\mu,\nu=1,\ldots,N$, is the Hessian matrix of $V$
evaluated at $\big(u,f(u)\big) \in \mathcal{K}$, and $F^J_r(u)$ is the matrix
of unknowns that we want to solve for. In the whole document, we adhere to the
practice of using different types of indices in order to indicate different
ranges of variation. Here, for example, $\mu,\nu,\rho,\ldots$ run from $1$ to
$N$; $r,s,t,\ldots$ run from $1$ to $K$; and $I,J,M,\ldots$ run from $K+1$ to
$N$. In the next section, we need to use more types of indices, but the idea
is the same.

 It is worth mentioning that similar equations to the ones above can be found
in classical mechanics anytime that local coordinates are used (the
coordinates $u$ in this work). For example, the force in such a case is
defined as $\partial V / \partial u^r$ and the chain rule can be used in a
similar way to what we do here. Note, however, that eq.~(\ref{eq:minimum2})
does not contain derivatives with respect to $u^r$, but to the constrained
coordinate $d^I$. This makes the approach slightly different and, indeed,
eq.~(\ref{eq:derderV}) would become trivial in the most common hard situation
tackled in the literature, where $\partial f^J / \partial u^r = 0$, $\forall
J, r$.

Since we are, by hypothesis, in a minimum of $V$ with respect to the
constrained coordinates $d$, the constrained sub-block $H_{JI}(u)$ of the
Hessian is a positive definite matrix, and therefore invertible. Hence, if we
multiply eq.~(\ref{eq:derderV}) by its inverse, denoted by $H^{IM}(u)$, sum
over $I$, exploit the fact that $H_{\mu\nu}(u)$ and $H^{\mu\nu}(u)$ are
symmetric, and conveniently rename the indices, we arrive at:
\begin{equation}
\label{eq:fsolved}
\frac{\partial f^I}{\partial u^r}(u) =: F^I_r(u) =
  - H^{IJ}(u) H_{Jr}(u) \ ,
\end{equation}
which, as promised, allows us to find the exact derivatives $(\partial f^I /
\partial u^r)(u)$ with the only knowledge of the Hessian of $V$ at the point
$\big(u,f(u)\big)$, and, upon introduction of the result in
eq.~(\ref{eq:derZ1bis}), also the derivatives along the constrained subspace
$\mathcal{K}$ of any physical observable $X(q)$.

As mentioned, several methods exist in the literature
\cite{Chr2005JCP,Hes2002JCP,Zho2000JCP} to perform MD simulations with
flexible constraints, however, none of them has tackled the calculation of
these derivatives, which are very basic objects presumably to be needed in
many future applications (see, e.g., refs.~\cite{Ech2006JCC2,Ech2011Sub2}). Of
course, it is always possible to compute derivatives using the simple and
straightforward method of finite differences. In this work, we use finite
differences as a way of validating the new, exact method and ensuring it is
error free.

The accuracy of the new algorithm is only limited by the accuracy with which
we can calculate the Hessian of $V$ at $\big(u,f(u)\big)$ and invert it; there
is no tunable parameter that we need to adjust for optimal accuracy, as in the
case of finite differences (see below and also Results and Discussion). This
makes a difference because, in classical force fields \cite{Pon2003APC} and
even in some quantum chemical methods (e.g., see chap.~10 of
\cite{Jen1998Book}), the Hessian can be calculated analytically, without the
need of finite differences.

Although no optimization of the numerical cost has been pursued in this work,
some remarks can be made about it, in comparison with the cost of the
finite-differences approach. In order to calculate the partial derivatives
$\partial Z/\partial u^s$ with respect to the unconstrained coordinates $u$
using finite differences, we need to:

\begin{enumerate}
\item Minimize $V(u,d)$ at fixed $u$ to find $f(u)$ (this step 
  is common with the new method introduced here).
\item Calculate $Z(u):=X\big(u,f(u)\big)$ (this step is
  common with the new method introduced here).
\item Choose a displacement $\Delta$ and minimize $V(\tilde{u},d)$ at the
  point $\tilde{u}:=(\tilde{u}^r)_{r=1}^K$, where $\tilde{u}^r=u^r + \Delta$
  if $r=s$ and $\tilde{u}^r=u^r$ if $r \neq s$. This yields $f(\tilde{u})$ at 
  a nearby point in $\mathcal{K}$ with $u^s$ displaced a quantity $\Delta$ and 
  the rest of unconstrained coordinates kept the same.
\item Calculate $Z(\tilde{u}):=X\big(\tilde{u},f(\tilde{u})\big)$.
\item Calculate
\begin{equation}
\label{eq:finite_differences}
\frac{\Delta Z}{\Delta u^s} := \frac{Z(\tilde{u}) - Z(u)}{\Delta}	
\end{equation}
  as the finite-difference approximation to the sought derivative
  $\partial Z/\partial u^s$ at the point $u$.
\end{enumerate}

Note that the third point of this finite-differences approach is essentially
a linear stability analysis. When strongly non-equilibrium points are
present, such as in the examples discussed in the last section, this
approximation fails and the fact that the new algorithm introduced in this
work uses only quantities defined \emph{at the point} $u$ becomes even more
important.

Now, assuming that we have a good enough guess for the parameter $\Delta$, the
cost of this procedure is dominated by the need to perform $K$ minimizations
of the function $V$, one in each of the directions corresponding to the
unconstrained coordinates $u^r$. If we denote by $N_\mathrm{it}$ the average
number of iterations needed for these minimizations to converge, and we define
$C_V$ and $C_{dV}$ as the numerical costs (in computer time) of computing $V$
and its first derivatives with respect to the constrained coordinates $d$,
respectively, we have that the average cost of calculating the sought
derivatives $\partial Z/\partial u^r$ using finite differences will be $K
N_\mathrm{it} (C_V + C_{dV})$ for local optimization methods such as the
steepest descent or the conjugate gradient, or $K N^\prime_\mathrm{it} C_V$
for Monte Carlo-based methods in which the derivatives of $V$ are not needed,
such as simulated annealing \cite{Pre2007Book}.

On the other hand, the new algorithm does not require the extra minimizations,
but it does require the calculation of the Hessian of $V$ with respect to the
internal coordinates (whose cost we call $C_H$), and the computation of the
inverse of its constrained sub-block, $H_{JI}$, applied to each one of the $K$
$L$-vectors $F_r$ in eqs.~(\ref{eq:derderV}) and~(\ref{eq:fsolved}), of cost
$C_{iH}$; resulting in a total cost of $C_H + K C_{iH}$.

The comparison between the two costs is not trivial and some remarks about it
must be made: First, one must notice that the different individual costs
involved, $C_V$, $C_{dV}$, $C_H$ and $C_{iH}$, are strongly dependent on the
characteristics (1) of the coordinates $q$ used and (2) of the function
$V(q)$. For example, if the coordinates $q$ are the Euclidean ones and the
function $V(q)$ is the potential energy of a molecular system as modeled by a
typical force field
\cite{Bro2009JCC,Jor1988JACS,Jor1996JACS,Pon2003APC,AMBER10,Pea1995CoPC}, the
most direct algorithms for calculating $V$ and its derivatives yield costs
$C_V$, $C_{dV}$ and $C_H$ which are of order $N^2$, $NK$ and $N^2$,
respectively \cite{Fre2002Book}. However, if more advanced long-range
techniques are used, such as the particle-particle particle-mesh (PPPM) method
\cite{Eas1974JCoP}, the fast multipole method \cite{Gre1987JCoP} or the
particle-mesh Ewald summation \cite{Dar1993JCP}, these costs can be reduced to
order $N\log N$ or even $N$ (for large $N$ and forgetting prefactors). Also,
as mentioned, if the coordinates used are not the Euclidean ones but some
internal coordinates such as the ones used in this work, these costs must
change in order to account for the transformation between the two. If force
fields are not used but, instead, $V(q)$ is the ground-state Born-Oppenheimer
energy as calculated using Hartree-Fock \cite{Ech2007MP}, then the most naive
implementations yield costs for $C_V$, $C_{dV}$ and $C_H$ which are of order
$N^4$ \cite{Jen1998Book}. The cost, $C_{iH}$, of calculating the inverse of
$H_{JI}$ applied to a vector $F_r$ can range from order $N$ to order $N^3$
depending on the sparsity of the matrix \cite{Pre2007Book}, which, in turn,
depends again on the coordinates used and on the structure of $V(q)$. Finally,
additional qualifications may complicate the comparison, such as the
architecture of the computers in which the algorithms are implemented,
parallelization issues, or the fact that, e.g., if we need the Hessian for a
different purpose in our simulation, such as the calculation of the
corresponding correcting term that appears both in the constrained stiff model
and in the Fixman potential \cite{Ech2006JCC2}, then the `only' computational
step we are adding is the inversion of a matrix.

Despite the complexity and problem-dependence of the cost assessment, it must
be stressed that, even in the cases in which the new algorithm turns out to be
more expensive than the alternatives, the fact that it is exact and
parameter-free might still make it the preferred choice in problems where high
accuracy is needed. Although a parameter-free structure does not
guarantee higher accuracy, in this case it does, since our method can be
identified as the proper limit of the finite-differences scheme when $\Delta
\to 0$. This is illustrated in Results and Discussion.

It is also worth remarking that the new method, as mentioned, is not needed to
perform MD simulations, which can be run without calculating any of the
derivatives tackled in this work \cite{Chr2005JCP,Hes2002JCP,Zho2000JCP}. Our
method is only needed when some observable in which these derivatives are
included (such as the aforementioned mass-metric tensor determinants) needs to
be computed. In such cases, the only two options to get to the final result
are either finite differences or our method, and the most convenient of the
two has to be chosen; even if its cost is a burden.

\subsection*{Application to Euclidean Coordinates of Molecules}

In this section, we will apply the general algorithm introduced above to
calculate the derivatives along the constrained subspace of the Euclidean
coordinates of molecular systems in a frame of reference (FoR) fixed in the
molecule. This problem has been faced by our group when trying to calculate
the correcting terms associated with mass-metric tensor determinants that
appear in the equilibrium probability density when flexible constraints are
imposed \cite{Ech2006JCC2,Ech2006JCC3,Ech2011Sub2}. More specifically, these
derivatives are needed to calculate the determinant of the induced mass-metric
tensor $g$ that appears in the constrained rigid model, according to the
formulae derived in ref.~\cite{Ech2006JCC3}.

In such a case, the system of interest is a set of $n$ mass points termed
\emph{atoms}. The three Euclidean coordinates of atom $\alpha$ in a FoR fixed
in the laboratory are denoted by $\vec{x}_\alpha$, and its mass by $m_\alpha$,
with $\alpha=1,\ldots,n$. However, when no explicit mention to the atom index
needs to be made, we will use $x := (x^\mu)_{\mu=1}^{N}$ to denote the
$N$-tuple of all the $N:=3n$ Euclidean coordinates of the system. The masses
$N$-tuple, $m := (m_\mu)_{\mu=1}^{N}$, in such a case, is formed by
consecutive groups of three identical masses, corresponding to each of the
atoms.

Apart from the Euclidean coordinates, one can also use a given set of
\emph{curvilinear coordinates} (also called sometimes \emph{general} or
\emph{generalized}), denoted by $q := (q^\mu)_{\mu=1}^N$, to describe the
system. Both the coordinates $x$ and $q$ parameterize the whole space
$\mathcal{W}$, and the transformation between the two sets and its inverse are
respectively given by
\begin{subequations}
\label{eq:change_x_to_q}
\begin{align}
& x^\mu = X^\mu(q) \ , \qquad \mu=1,\ldots,N \ ,
  \label{eq:change_x_to_q_a} \\
& q^\mu = Q^\mu(x) \ , \qquad \mu=1,\ldots,N \ .
  \label{eq:change_x_to_q_b}
\end{align}
\end{subequations}

We will additionally assume that, for the points of interest, this is a 
\emph{proper} change of coordinates, i.e., that the \emph{Jacobian matrix}
\begin{equation}
\label{eq:Jacobian_x_to_q}
J^\mu_\nu := \frac{\partial X^\mu(q)}{\partial q^\nu}
\end{equation}
has non-zero determinant.

\begin{figure}[!ht]
\begin{center}
\includegraphics[scale=0.4]{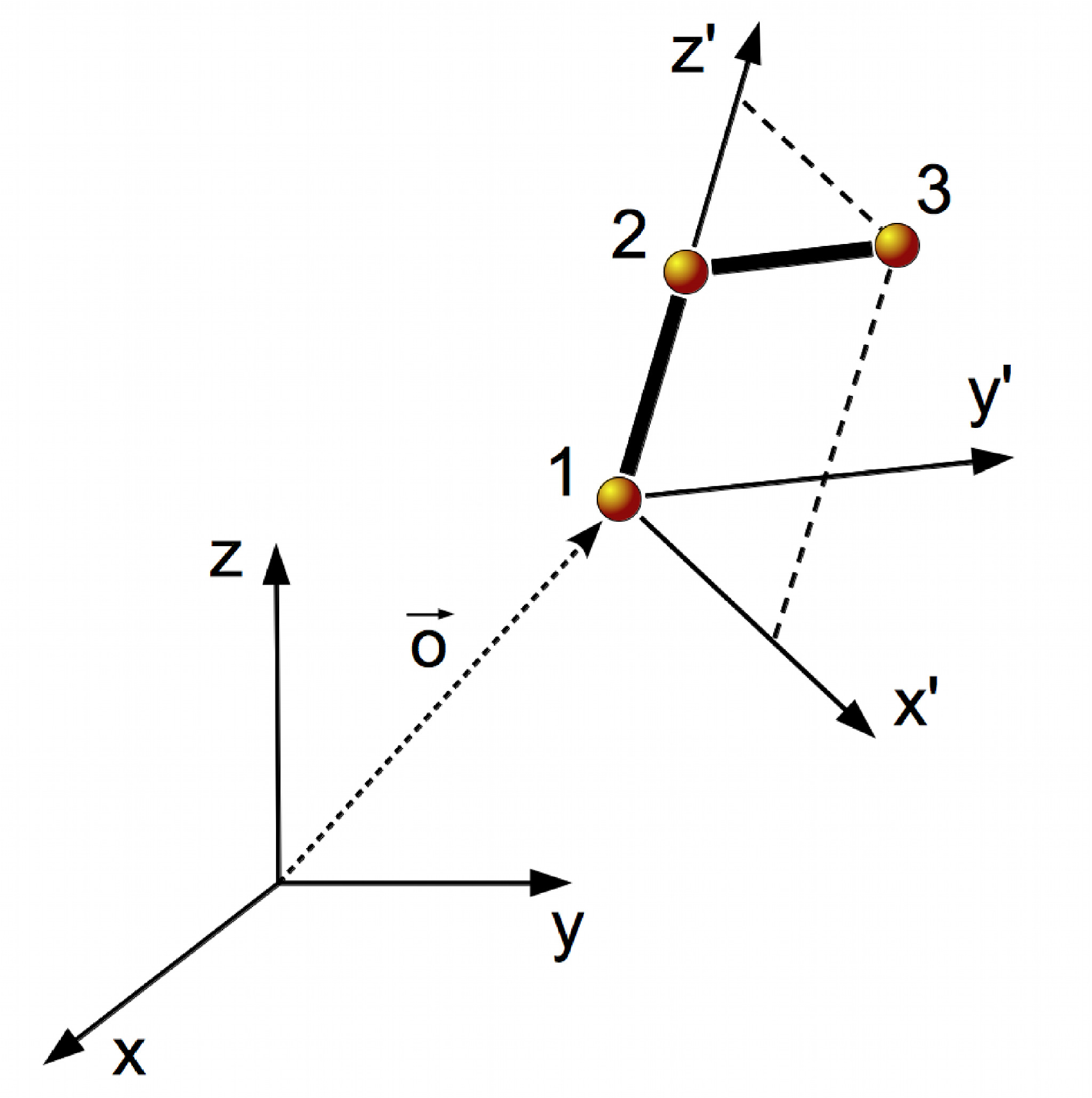}
\end{center}
\caption{
{\bf Definition of the frame of reference fixed in the system.}
}
\label{fig:axes_fixed}
\end{figure}

Now, we define a particular FoR \emph{fixed in the system} to perform some of
the calculations. To this end, we select three atoms (denoted by 1, 2 and 3)
in such a way that $\vec{o}$, the position in the FoR of the laboratory of the
origin of the FoR fixed in the system, is the Euclidean position of atom 1
(i.e., $\vec{o}:=\vec{x}_{1}$). The orientation of the FoR
$(x^{\,\prime},y^{\,\prime},z^{\,\prime})$ fixed in the system is chosen such
that atom 2 lies in the positive half of the $z^{\,\prime}$-axis, and atom 3
is contained in the $(x^{\,\prime},z^{\,\prime})$-plane, with projection on
the positive half of the $x^{\,\prime}$-axis (see fig.~\ref{fig:axes_fixed}).
The position of any given atom $\alpha$ in the new FoR fixed in the system is
denoted by $\vec{x}_{\alpha}^{\,\prime}$. Also, let $E(\phi,\theta,\psi)$ be
the Euler rotation matrix (in the ZYZ convention) that takes a free 3-vector
of primed components, $\vec{a}^{\,\prime}$, to the FoR fixed in the
laboratory, i.e., $\vec{a}=E(\phi,\theta,\psi)\,\vec{a}^{\,\prime}$
\cite{Gol2002Book}.

Although the aforementioned curvilinear coordinates $q$ are a priori general,
it is very common to take into account the fact that the typical potential
energy functions of molecular systems in absence of external fields do not
depend on $\vec{o}^T:=(o_x,o_y,o_z)$ nor on the angles $(\phi,\theta,\psi)$,
and to consequently choose a set of curvilinear coordinates split into $q =
(e,r)$, where the first six are these \emph{external coordinates}, $e :=
(e^A)_{A=1}^6 = (o_x,o_y,o_z,\phi,\theta,\psi)$. As we mentioned before,
$\vec{o}:=(o_x,o_y,o_z)$ describes the overall position of the system with
respect to the FoR fixed in the laboratory, and its overall orientation is
specified by the angles $(\phi,\theta,\psi)$. The remaining $N-6$ coordinates
$r := (r^a)_{a=7}^N$ are called \emph{internal coordinates} and determine
the positions of the atoms in the FoR fixed in the system
\cite{Ech2007CoP,Ech2006JCC1}. They parameterize what we shall call the
\emph{internal subspace} or \emph{conformational space}, denoted by
$\mathcal{I}$, and the coordinates $e$ parameterize the \emph{external
subspace}, denoted by $\mathcal{E}$; consequently splitting the whole space as
$\mathcal{W} = \mathcal{E} \times \mathcal{I}$ (denoting by $\times$ the
Cartesian product of sets).

The position, $\vec{x}_\alpha^{\,\prime}$, of any given atom $\alpha$ in the
axes fixed in the system is a function, $\vec{X}^{\,\prime}_\alpha(r)$, of
only the internal coordinates, $r$, and the transformation from the Euclidean
coordinates $x$ to the curvilinear coordinates $q$
in~(\ref{eq:change_x_to_q_a}) may be written more explicitly as follows:
\begin{equation}
\label{eq:change_x_to_q_2}
\vec{x}_\alpha = \vec{X}_\alpha(q) = \vec{o} + E(\phi,\theta,\psi)\,
 \vec{X}^{\,\prime}_\alpha(r) \ .
\end{equation}

Although general constraints affecting all the coordinates $q$ [like those
in~(\ref{eq:constraints1})] can be imposed on the system, the already
mentioned property of invariance of the potential energy function under
changes of the external coordinates, $e$, together with the fact that the
potential energy can be regarded as `producing' the constraints
\cite{Ech2006JCC2}, make physically frequent the use of constraints
involving only the internal coordinates, $r$:
\begin{equation}
\label{eq:constraints3}
h^I(r) = 0  \ , \quad I=K+1,\ldots,N \ .
\end{equation}

Under the common assumptions in the Introduction, these constraints allow us
to split the internal coordinates as $r=(s,d)$, where the first $M:=K-6=N-L-6$
ones, $s := (s^i)_{i=6+1}^K$, are called \emph{unconstrained internal
coordinates} and parameterize the \emph{internal constrained subspace},
denoted by $\Sigma$. The last $L:=N-K$ ones, $d := (d^I)_{I=K+1}^N$,
correspond to the \emph{constrained coordinates} in the Introduction and are
called. The external coordinates, $e$, together with the unconstrained
internal coordinates, $s$, constitute the set of all \emph{unconstrained
coordinates} of the system, $u=(e,s)$, which parameterize the constrained
subspace $\mathcal{K}$, being $\mathcal{K} = \mathcal{E} \times \Sigma$.

In this situation, the constraints in eq.~(\ref{eq:constraints3}) are
equivalent to
\begin{equation}
\label{eq:constraints4}
d^I = f^I(s)  \ , \quad I=K+1,\ldots,N \ ,
\end{equation}
and the functions $f^I(s)$ are defined as taking the values of the coordinates
$d^I$ that minimize the potential energy with respect to all $d^I$ at fixed
$s$ \cite{Ech2006JCC2,Ech2011Sub2,Zho2000JCP}.

Finally, if these constraints are used, together
with~(\ref{eq:change_x_to_q_2}), the Euclidean position of any atom in the
constrained case may be parameterized with the set of all unconstrained
coordinates, $u$, as follows:
\begin{eqnarray}
\label{eq:transf_constrained}
\vec{x}_\alpha &=& \vec{Z}_\alpha(u) := \vec{X}_\alpha\big(e,s,f(s)\big)
\nonumber \\
&=& \vec{o} + E(\phi,\theta,\psi) \vec{X}^{\,\prime}_\alpha\big(s,f(s)\big)
 \nonumber \\
&=:& \vec{o} + E(\phi,\theta,\psi) \vec{Z}^{\,\prime}_\alpha(s) \ ,
\end{eqnarray}
where the name of the transformation functions has been changed from $X$ to
$Z$, and from $X^{\,\prime}$ to $Z^{\,\prime}$, in order to emphasize that the
dependence on the coordinates is different between the two cases.

In order to calculate the derivatives along $\Sigma$ of the primed atoms
positions, $\vec{Z}^{\,\prime}_\alpha$, with respect to the unconstrained
internal coordinates $s$ (needed, for example, in eq.~(28) of
ref.~\cite{Ech2006JCC3} to compute the determinant of the induced mass-metric
tensor $g$), we first differentiate with respect to $s^i$ in
$\vec{Z}^{\,\prime}_\alpha(s) := \vec{X}^{\,\prime}_\alpha \big(s,f(s)\big)$,
arriving to the analogue to eq.~(\ref{eq:derZ1bis}):
\begin{equation}
\label{eq:der_xp}
\frac{\partial \vec{Z}^{\,\prime}_\alpha}{\partial s^i}(s) =
\frac{\partial \vec{X}^{\,\prime}_\alpha}{\partial s^i} \big(s,f(s)\big) +
\frac{\partial \vec{X}^{\,\prime}_\alpha}{\partial d^I}
\big(s,f(s)\big) \frac{\partial f^I}{\partial s^i}(s)  \ .
\end{equation}

Now, the derivatives $(\partial f^I / \partial s^i)(s)$ can be calculated
using the general algorithm introduced in the previous section
simply noticing that, in this case, $V$ is precisely the potential energy of
the system. Therefore, the only objects that remain to be computed are the
derivatives $\partial \vec{X}^{\,\prime}_\alpha/\partial s^i$ and $\partial
\vec{X}^{\,\prime}_\alpha/\partial d^I$, which can be known analytically (they
are \emph{geometrical} [or \emph{kinematical}] objects, i.e., they do not
depend of the potential energy). We now turn to the derivation of an explicit
algorithm for finding them and thus completing the calculation that is the
objective of this section.

In the supplementary material of ref.~\cite{Ech2006JCC3}, we give a
detailed and explicit way for expressing any `primed' vector
$\vec{X}^{\,\prime}_\alpha$ as a function of all the internal coordinates, in
the particular coordination scheme known as SASMIC \cite{Ech2006JCC1}. We
could take the final expression there [eq.~(5)] and explicitly perform the
partial derivatives, however, we shall follow a different approach that is
both more straightforward and applicable to a larger family of Z-matrix-like
schemes for defining internal coordinates.

In non-redundant internal coordinates schemes, whether they are defined as in
ref.~\cite{Ech2006JCC1} or not, each atom is commonly regarded as being
incrementally added to the growing molecule for its coordination. This means
that the position of the $(\alpha > 3)$-th atom in the body-fixed axes is
uniquely specified by the values of three internal coordinates that are
defined with respect to the positions of three other atoms with indices
$\beta(\alpha), \delta(\alpha), \gamma(\alpha) < \alpha$. This is a very
convenient practice, and we will assume that we are dealing with a scheme that
adheres to it.

Normally, the first of the three internal coordinates used to position
atom $\alpha$ is the length of the vector joining $\alpha$ and
$\beta(\alpha)$. Atom $\beta(\alpha)$ is commonly chosen to be covalently
attached to $\alpha$ and, then, the length of this vector is naturally termed
\emph{bond length}, and denoted by $b_\alpha$. A given function
$\beta(\alpha)$ embodies the protocol used for defining this atom,
$\beta(\alpha)$, to which each `new' atom $\alpha$ is (mathematically)
attached; a superindex, as in $\beta^p(\alpha)$, indicates composition of
functions, and the iteration of such compositions allows us to trace a
single-branched chain of atoms that takes from atom $\alpha$ to atom 1, at the
origin of the `primed' axes. If $N_\alpha$ is a number such that
$\beta^{N_\alpha}(\alpha) = 1$, this chain is given by the following set:
\begin{equation}
\label{eq:betaalphaset}
\mathcal{B}_\alpha := \{\alpha=\beta^0(\alpha),\beta(\alpha), \beta^2(\alpha), 
 \ldots, \beta^{N_\alpha-3}(\alpha), 3, 2, 1\} \ .
\end{equation}

It is clear that, if we now change a given bond length $b_\varepsilon$
associated to atom $\varepsilon$, atom $\alpha$ will move if $\varepsilon \in
\mathcal{B}_\alpha$; simply because atom $\varepsilon$ \emph{will} move and
$\alpha$ has been positioned in reference to atom $\varepsilon$'s position.
Thus, if we define
\begin{equation}
\label{eq:Ralphabeta}
\vec{R}_{\alpha\beta} := \vec{X}^{\,\prime}_\beta - \vec{X}^{\,\prime}_\alpha
 \ ,
\end{equation}
for any $\alpha$, $\beta$, and accordingly denote by
$\hat{R}_{\beta(\varepsilon)\varepsilon}$ the unitary vector in the `primed'
FoR that points from atom $\beta(\varepsilon)$ to atom $\varepsilon$, a change
in the bond length associated to $\varepsilon$ from $b_\varepsilon$ to
$b_\varepsilon + db$ (while keeping the rest of the internal coordinates
constant) will translate all atoms $\alpha$ such that $\varepsilon \in
\mathcal{B}_\alpha$ a distance $db$ along
$\hat{R}_{\beta(\varepsilon)\varepsilon}$, having
\begin{equation}
\label{eq:dxp_dr1pre}
\vec{X}^{\,\prime}_{\alpha}(b_\varepsilon + db) = 
 \vec{X}^{\,\prime}_{\alpha}(b_\varepsilon) +
 \hat{R}_{\beta(\varepsilon)\varepsilon} db \ ,	
\end{equation}
and hence
\begin{equation}
\label{eq:dxp_dr1}
 \frac{\partial \vec{X}^{\,\prime}_{\alpha}}{\partial b_\varepsilon} := 
 \lim_{db \to 0} \frac{\vec{X}^{\,\prime}_{\alpha}(b_\varepsilon + db) -
 \vec{X}^{\,\prime}_{\alpha}(b_\varepsilon)}{db} =
 \hat{R}_{\beta(\varepsilon)\varepsilon} \ .
\end{equation}

\begin{figure}[!ht]
\begin{center}
\includegraphics[scale=0.4]{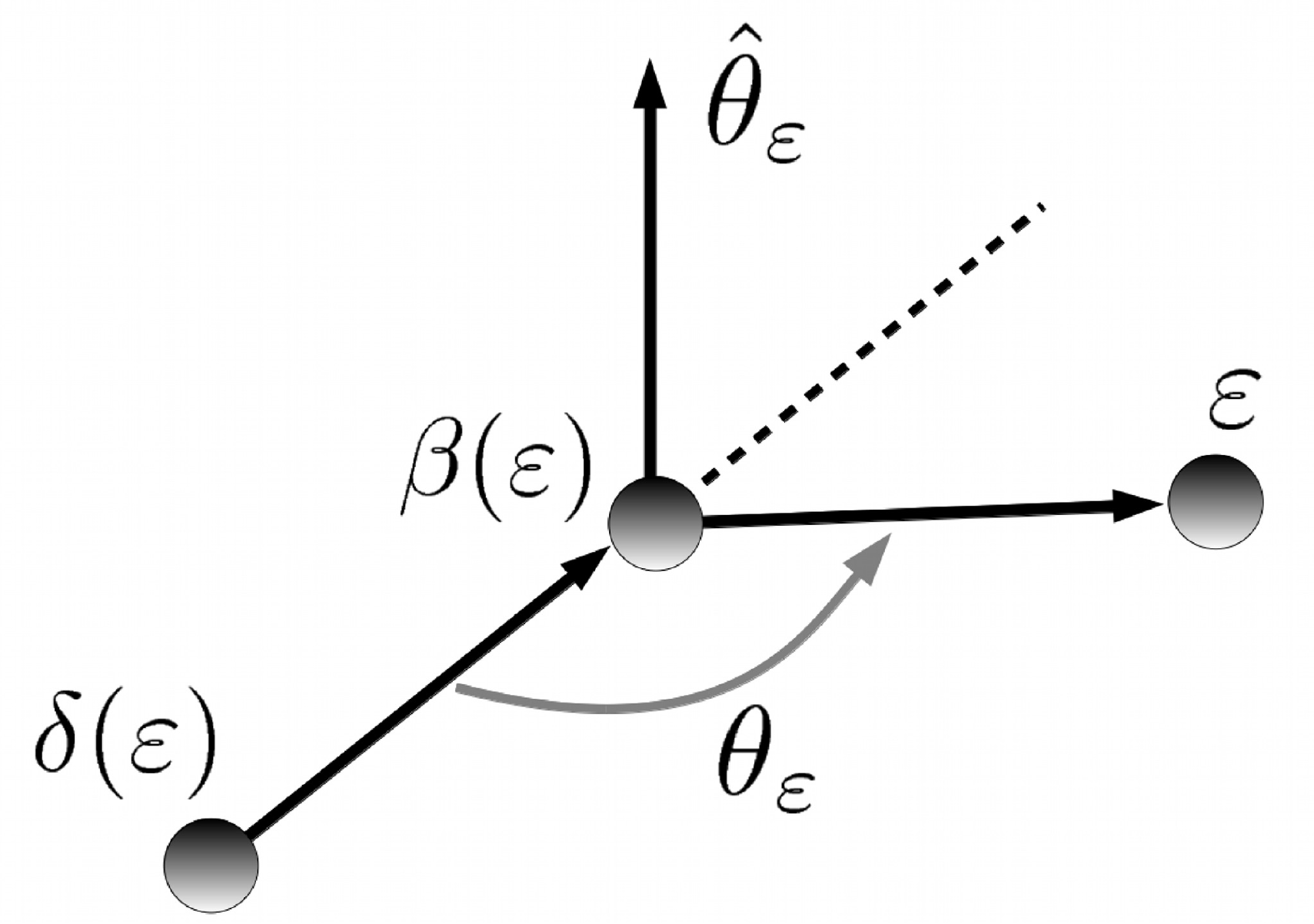}
\end{center}
\caption{
{\bf Rotation associated to a change in a bond angle.} Definition of the 
 \emph{bond angle} $\theta_\varepsilon$, associated to atom $\varepsilon$,
 and the unitary vector $\hat{\theta}_\varepsilon$ corresponding to the
 direction around which all atoms $\alpha$ with chains $\mathcal{B}_\alpha$
 containing $\varepsilon$ rotate if $\theta_\varepsilon$ is varied while
 the rest of internal coordinates are kept constant.
}
\label{fig:rotation_bond_angle}
\end{figure}

The second internal coordinate, after $b_\alpha$, that is typically defined to
position atom $\alpha$ with respect to the `already positioned' part of the
molecule is a so-called \emph{bond angle} $\theta_\alpha$. To define this
angle, we need an additional atom associated with $\alpha$, which we could
denote by $\delta(\alpha)$. Although one can in principle think of the
possibility of using different atoms $\beta_\theta(\alpha)$ and
$\delta_\theta(\alpha)$ to define the bond angle than the one used to define
the bond length, the common practice in the literature is to use the same
three atoms $\beta(\alpha)$, $\delta(\alpha)$ and $\gamma(\alpha)$, to define
the three internal coordinates associated to $\alpha$. This is also the choice
in the SASMIC scheme and the one in this work. The angle $\theta_\alpha$ is
thus defined as 180$^o$ minus the angle formed between the vectors
$\vec{R}_{\delta(\alpha)\beta(\alpha)}$ and $\vec{R}_{\beta(\alpha)\alpha}$
(see fig.~\ref{fig:rotation_bond_angle}).

Now, the reasoning is the same as in the case of the derivative with respect
to $b_\varepsilon$: For every atom $\varepsilon > 2$ that is the `tip' of the
bond angle $\theta_\varepsilon$, the changes in this angle (keeping the rest
of internal coordinates constant) will move atom $\varepsilon$ and therefore
all atoms $\alpha$ that contain atom $\varepsilon$ in the chain
$\mathcal{B}_\alpha$ that links them to atom 1.

If we now look at fig.~\ref{fig:rotation_bond_angle}, we see that a change
from $\theta_\varepsilon$ to $\theta_\varepsilon + d\theta$ amounts to rotate
all atoms $\alpha$ that contain $\varepsilon$ in their chain to the origin an
angle $d\theta$ around the unitary vector $\hat{\theta}_\varepsilon$, which is
defined by
\begin{equation}
\label{eq:theta_varepsilon}
	\hat{\theta}_\varepsilon :=
	 \frac{\vec{R}_{\beta(\varepsilon)\varepsilon} \times
	       \vec{R}_{\delta(\varepsilon)\beta(\varepsilon)}}
	      {|\vec{R}_{\beta(\varepsilon)\varepsilon} \times
		   \vec{R}_{\delta(\varepsilon)\beta(\varepsilon)}|} \ .
\end{equation}

The result, $\vec{v}_\mathrm{rot}$, of rotating a vector $\vec{v}$ around the
direction given by the unitary vector $\hat{\theta}$ an amount $\theta$ is
given by the well-known \emph{Rodrigues' rotation formula}
\cite{Bel2009Web,Gol2002Book}:
\begin{equation}
\label{eq:rodrigues}
	\vec{v}_\mathrm{rot} = \vec{v}\cos\theta + (\hat{\theta} \times \vec{v})
	  \sin\theta + \hat{\theta}(\hat{\theta} \cdot \vec{v})(1-\cos\theta) \ .
\end{equation}

However, notice that, in order to define a rotation, it is not enough to
specify the angle $\theta$ and the rotation axis $\hat{\theta}$, but one
additionally needs to specify a fixed point (which can actually be any of the
points in a fixed line in the direction of $\hat{\theta}$). Therefore, the
above expression is only correct for either `free' vectors $\vec{v}$ (i.e.,
those that are not associated to a given point in space), or for vectors
$\vec{v}$ whose starting point lies in the aforementioned fixed line.

The fixed point for the rotation we are interested in can be chosen to be 
$\beta(\varepsilon)$ and, using eq.~(\ref{eq:rodrigues}), we have that
\begin{eqnarray}
\label{eq:dr_dtheta1}
\vec{R}_{\beta(\varepsilon)\alpha}(\theta_\varepsilon + d\theta) & = &
 \vec{R}_{\beta(\varepsilon)\alpha}(\theta_\varepsilon)\cos d\theta
 \nonumber \\
 & & \mbox{} + \big[\hat{\theta}_\varepsilon \times 
 \vec{R}_{\beta(\varepsilon)\alpha}(\theta_\varepsilon) \big] \sin d\theta
 \\
 & & \mbox{} +  \hat{\theta}_\varepsilon \big[ \hat{\theta}_\varepsilon \cdot 
 \vec{R}_{\beta(\varepsilon)\alpha}(\theta_\varepsilon) \big]
   (1 - \cos d\theta)\ . \nonumber
\end{eqnarray}

Then, keeping the terms up to first order in $d\theta$, we can easily
compute the derivative:
\begin{equation}
\label{eq:dr_dtheta2}
\frac{\partial \vec{R}_{\beta(\varepsilon)\alpha}}
     {\partial \theta_\varepsilon} =
   \hat{\theta}_\varepsilon \times 
   \vec{R}_{\beta(\varepsilon)\alpha} \ ,
\end{equation}
which, since a variation of $\theta_\varepsilon$ does not move atom
$\beta(\varepsilon)$ (i.e. $\partial \vec{X}^{\,\prime}_{\beta(\varepsilon)}
/ \partial \theta_\varepsilon = 0$), allows us to conclude that
\begin{equation}
\label{eq:dxp_dtheta1}
	\frac{\partial \vec{X}^{\,\prime}_{\alpha}}
	     {\partial \theta_\varepsilon} =
	\frac{\partial }{\partial \theta_\varepsilon}\left(
 \vec{X}^{\,\prime}_{\beta(\varepsilon)} + \vec{R}_{\beta(\varepsilon)\alpha}
   \right) = 
\frac{\partial \vec{R}_{\beta(\varepsilon)\alpha}}
     {\partial \theta_\varepsilon} =
   \hat{\theta}_\varepsilon \times 
   \vec{R}_{\beta(\varepsilon)\alpha} \ ,
\end{equation}
if $\varepsilon \in \mathcal{B}_\alpha$.

\begin{figure}[!ht]
\begin{center}
\includegraphics[scale=0.4]{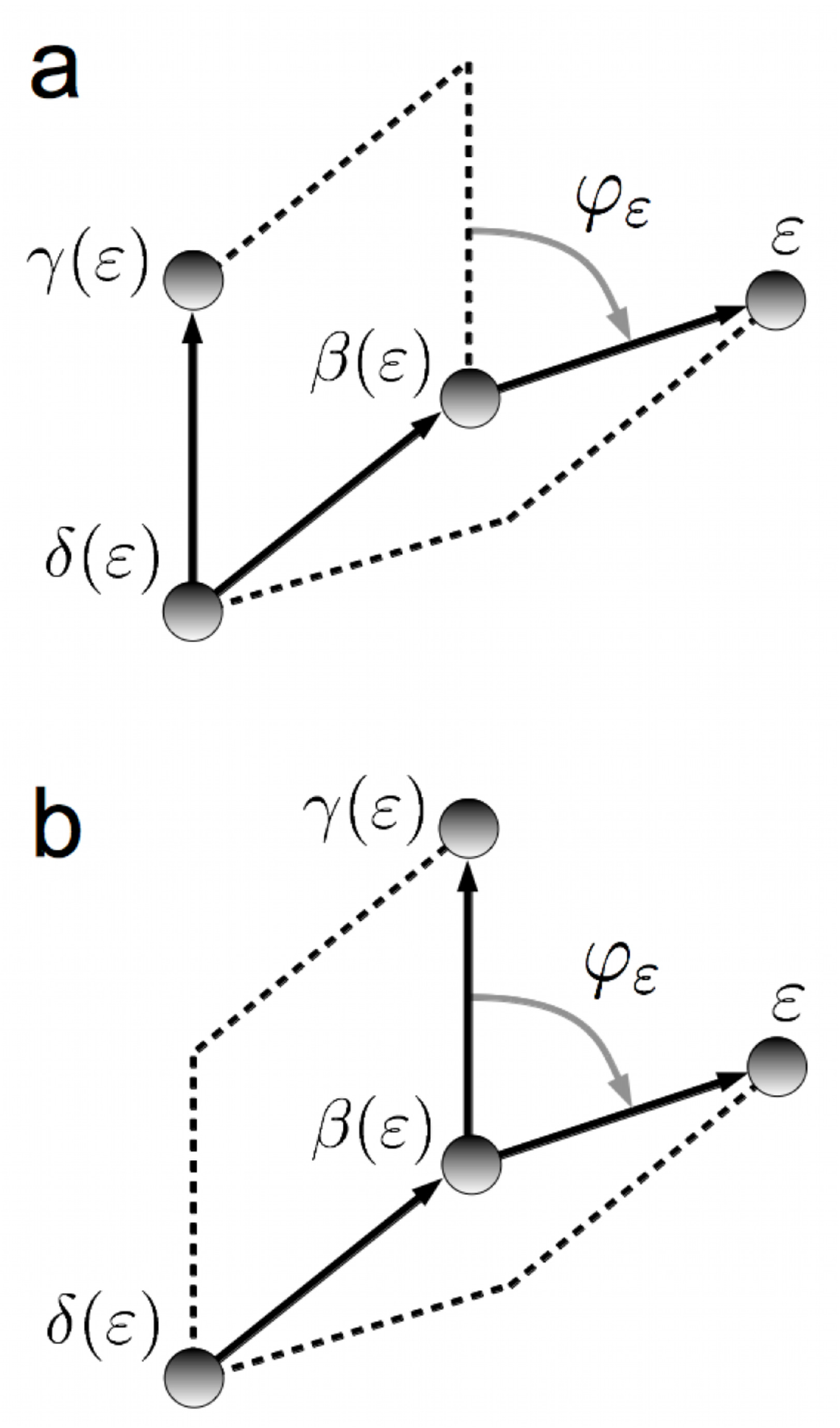}
\end{center}
\caption{
{\bf Rotation associated to a change in a dihedral angle.} Definition of the 
 \emph{dihedral angle} $\varphi_\varepsilon$, associated to atom 
 $\varepsilon$. The positive sense of rotation is indicated in the figure, and 
 we can distinguish between two situations regarding covalent connectivity:
 \textbf{a)} \emph{principal dihedral angle}, and \textbf{b)} \emph{phase
  dihedral angle} (see ref.~\cite{Ech2006JCC1}).
}
\label{fig:rotation_dihedral_angle}
\end{figure}

The third and last internal coordinate that is usually defined to position
atom $\alpha$ is a so-called \emph{dihedral angle} $\varphi_\alpha$. To define
this angle, we need a third atom associated with $\alpha$, which we could
denote by $\gamma(\alpha)$. The angle $\varphi_\alpha$ is thus defined as the
oriented angle formed between the plane containing atoms $\beta(\alpha)$,
$\delta(\alpha)$ and $\gamma(\alpha)$ and the plane containing atoms $\alpha$,
$\beta(\alpha)$ and $\delta(\alpha)$. The positive sense of $\varphi_\alpha$
is the one indicated in fig.~\ref{fig:rotation_dihedral_angle}, and, although
it is common to find two different covalent arrangements of the four atoms
$\alpha$, $\beta(\alpha)$, $\delta(\alpha)$ and $\gamma(\alpha)$, termed
\emph{principal} and \emph{phase} dihedral angles, respectively
\cite{Ech2006JCC1}, this does not affect the mathematical definition of
$\varphi_\alpha$ given in this paragraph, nor the subsequent calculations.

Regarding the derivative of the `primed' position of atom $\alpha$ with
respect to a given $\varphi_\varepsilon$, the only difference with the bond
angle case is that, now, the rotation is performed around the direction given
by the unitary vector $\hat{R}_{\delta(\varepsilon)\beta(\varepsilon)}$ (see
fig.~\ref{fig:rotation_dihedral_angle}). The fixed point can be again chosen
as $\beta(\varepsilon)$, and eq.~(\ref{eq:dr_dtheta2}) (changing
$\theta_\varepsilon$ by $\varphi_\varepsilon$ and $\hat{\theta}_\varepsilon$
by $\hat{R}_{\delta(\varepsilon)\beta(\varepsilon)}$), as well as the fact
that changes in $\varphi_\varepsilon$ do not move atom $\beta(\varepsilon)$,
still hold. Therefore,
\begin{equation}
\label{eq:dxp_dphi1}
	\frac{\partial \vec{X}^{\,\prime}_{\alpha}}
	     {\partial \varphi_\varepsilon} =
   \hat{R}_{\delta(\varepsilon)\beta(\varepsilon)} \times 
   \vec{R}_{\beta(\varepsilon)\alpha} \ ,
\end{equation}
if $\varepsilon \in \mathcal{B}_\alpha$.

In order to decide whether or not atom $\alpha$ will move upon changes in
internal coordinates associated to atoms $\varepsilon$ that \emph{do not}
belong to $\mathcal{B}_\alpha$ we must first finish the story about internal
coordinates definition. Since the argument above to show that $\alpha$ moved
when $\varepsilon \in \mathcal{B}_\alpha$ was that $\varepsilon$ \emph{itself}
moved and it was used to position $\alpha$, we must ask
\begin{enumerate}
\item whether or not there can be atoms that are also used to position 
  $\alpha$ but that do not belong to $\mathcal{B}_\alpha$, and
\item what happens when we change the internal coordinates associated to
  them.
\end{enumerate}

\begin{figure}[!ht]
\begin{center}
\includegraphics[scale=0.4]{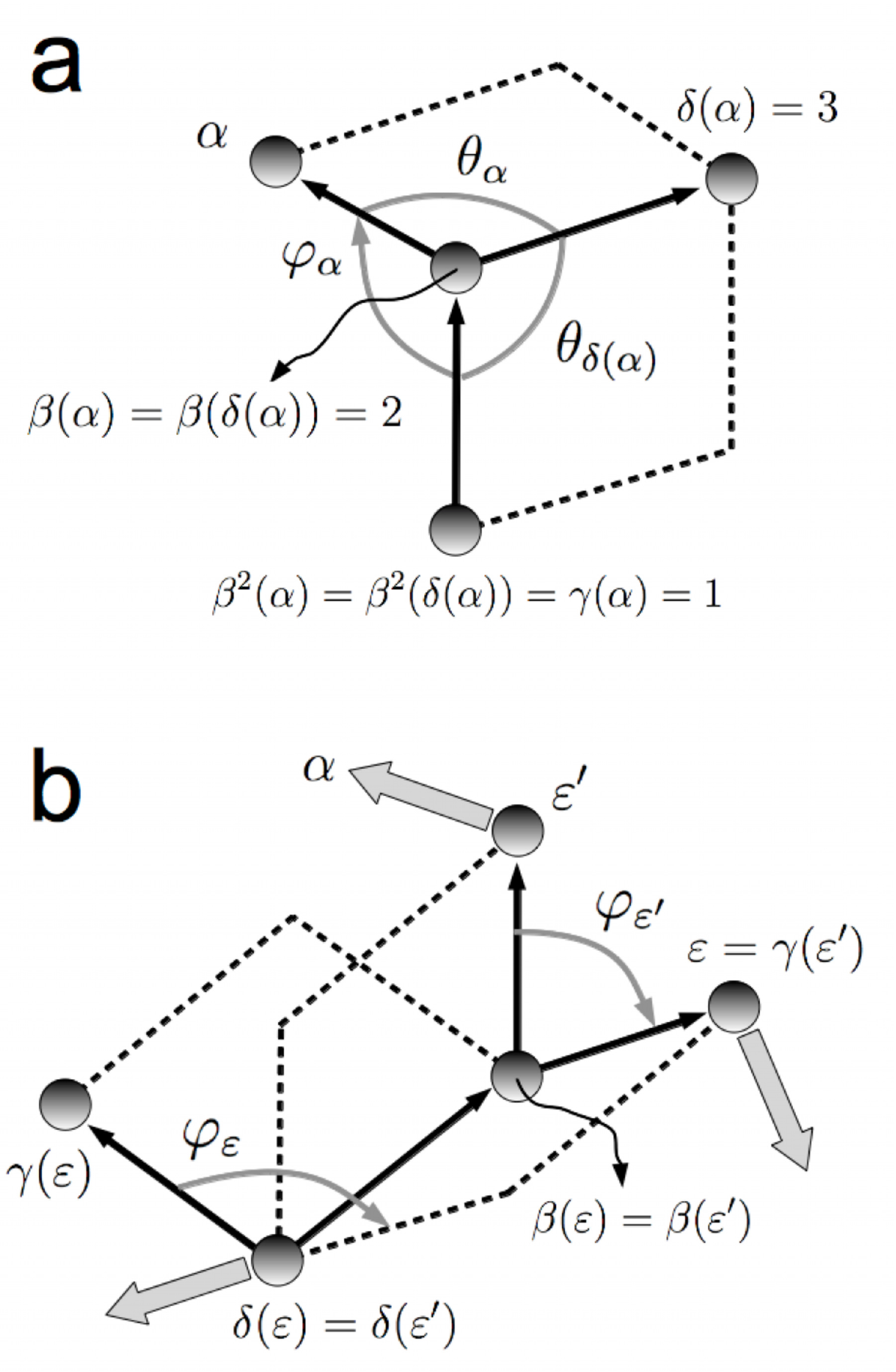}
\end{center}
\caption{
{\bf Special cases.} Special cases of atoms that do not
 belong to the chain $\mathcal{B}_\alpha$ connecting $\alpha$ to atom 1, but
 that are nevertheless used to position $\alpha$.
}
\label{fig:special_cases}
\end{figure}

The answers to these two questions depend on the particular scheme used to
define the internal coordinates, and we will tackle them referring to the
SASMIC scheme \cite{Ech2006JCC1}, which is the one used in this work:
According to the SASMIC rules, there are only two situations in which an
atom $\varepsilon \notin \mathcal{B}_\alpha$ can be used to position atom
$\alpha$, and they are depicted in fig.~\ref{fig:special_cases}.

The first case, in fig.~\ref{fig:special_cases}a, attains only the first atoms
of the molecule. Typically, atom 1 is not a first-row atom, but a hydrogen
(such is the case of the three molecules studied, for example, in Results and
Discussion). Hence, after positioning atoms 2 and 3, which are typically
first-row, it is more representative to choose atom 3 as $\delta(\alpha)$ and
atom 1 as $\gamma(\alpha)$ when positioning the rest of the atoms $\alpha$
attached to atom 2. This makes $1 = \beta^2(\alpha) \neq \delta(\alpha)$ and
hence $\delta(\alpha)$ qualifies as an atom that is used to position $\alpha$
but which is not included in the chain $\mathcal{B}_\alpha$.

The second case, in fig.~\ref{fig:special_cases}b, corresponds to the
situation in which the molecule divides in two branches, and it can happen all
along its chemical structure. If atom $\varepsilon$ is the atom that defines
the only principal dihedral over the bond connecting $\delta(\varepsilon)$ and
$\beta(\varepsilon)$ (in the SASMIC scheme, only one principal dihedral can be
defined on a given bond \cite{Ech2006JCC1}), and atom $\alpha$ belongs to a
different branch than the one beginning in $\varepsilon$ (the branches are
indicated with grey broad arrows), then the starting atom $\varepsilon^\prime$
of the branch to which $\alpha$ belongs ($\varepsilon^\prime$ can be $\alpha$
itself) must be positioned using a phase dihedral in which
$\varepsilon=\gamma(\varepsilon^\prime)$. Thus, $\varepsilon$ is an atom that
is used to position $\alpha$, but which does not belong to the chain
$\mathcal{B}_\alpha$ connecting $\alpha$ to atom 1.

In principle, any change in the internal coordinates of atom 
$\delta(\alpha)$, in the first case, or in those of atom $\varepsilon$, in
the second case, may move atom $\alpha$, however, due to the geometrical
characteristics of the internal coordinates, this is not the case.

For example, it is easy to see that, in the case depicted in
fig.~\ref{fig:special_cases}a, a variation of the bond length $b_\varepsilon$
(denoting $\varepsilon:=\delta(\alpha)$) does not move atom $\alpha$.
Regarding the angles, the dihedral $\varphi_\varepsilon$ is not defined
because $\varepsilon=3$, and a change in $\theta_\varepsilon$ can be seen to
rotate atom $\alpha$ with fixed point $\beta(\varepsilon)=\beta(\alpha)$ and
around the axis given exactly by $\hat{\theta}_\varepsilon$ as defined in
eq.~(\ref{eq:theta_varepsilon}). (It is not trivial to see that this motion
keeps all the rest of internal coordinates constant, specially the phase
dihedral $\varphi_\alpha$. The authors found it helpful to imagine that atoms
1, 2 and 3 lie in the plane of the paper, with $\hat{\theta}_\varepsilon$ and
$\alpha$ coming out of it towards the reader; the first orthogonally and the
second not.) Therefore, the derivative of the Euclidean position of atom
$\alpha$ with respect to $\theta_\varepsilon$ is also given by
eq.~(\ref{eq:dxp_dtheta1}) in this special case.

In the situation shown in fig.~\ref{fig:special_cases}b, one can see that
neither a change in $b_\varepsilon$ nor in $\theta_\varepsilon$ move atom
$\varepsilon^\prime$ nor $\alpha$. However, if we change
$\varphi_\varepsilon$, we need to move atom $\varepsilon^\prime$ if we want to
keep $\varphi_{\varepsilon^\prime}$ constant. Therefore, atom $\alpha$ moves
in such a case and it does so by rotating with the same fixed point
$\beta(\varepsilon)$ and the same axis
$\hat{R}_{\delta(\varepsilon)\beta(\varepsilon)}$ as in the simpler cases
depicted in fig.~\ref{fig:rotation_dihedral_angle}. This means that, again, we
can calculate the sought derivative using the already justified
eq.~(\ref{eq:dxp_dphi1}).

In summary, only changes in bond lengths associated to atoms
$\varepsilon \in \mathcal{B}_\alpha$ affect the position of atom $\alpha$:
\begin{equation}
\label{eq:dxp_dr}
\frac{\partial \vec{X}^{\,\prime}_{\alpha}}{\partial b_\varepsilon} =
\begin{cases}
 \hat{R}_{\beta(\varepsilon)\varepsilon}
  & \mathrm{if} \ \varepsilon \in \mathcal{B}_\alpha \\
0 & \mathrm{if} \ \varepsilon \notin \mathcal{B}_\alpha \\
\end{cases}\ ;
\end{equation}
changes both in bond angles associated to atoms
$\varepsilon \in \mathcal{B}_\alpha$ and to $\delta(\alpha)$
in fig.~\ref{fig:special_cases}a affect the position of atom $\alpha$:
\begin{equation}
\label{eq:dxp_dtheta}
\frac{\partial \vec{X}^{\,\prime}_{\alpha}}
     {\partial \theta_\varepsilon} =
\begin{cases}
\hat{\theta}_\varepsilon \times 
   \vec{R}_{\beta(\varepsilon)\alpha}
  & \mathrm{if} \ \varepsilon \in \mathcal{B}_\alpha \ , 
	\\ & \ \mathrm{or} \ 
	\big[\beta(\alpha)=2 , \ \delta(\alpha)=3)\big] \\
0 & \mathrm{otherwise} \\
\end{cases}\ ;
\end{equation}
and changes both in dihedral angles associated to atoms $\varepsilon \in
\mathcal{B}_\alpha$ and to those that define the principal dihedral at a
branching point that leads to atom $\alpha$ (see
fig.~\ref{fig:special_cases}b) can affect the position of atom $\alpha$:
\begin{equation}
\label{eq:dxp_dphi}
\frac{\partial \vec{X}^{\,\prime}_{\alpha}}
     {\partial \varphi_\varepsilon} =
\begin{cases}
\hat{R}_{\delta(\varepsilon)\beta(\varepsilon)} \times 
   \vec{R}_{\beta(\varepsilon)\alpha}
  & \mathrm{if} \ \varepsilon \in \mathcal{B}_\alpha \ ,
	\\ & \ \mathrm{or} \ 
	\big[\varphi_\varepsilon \ \mathrm{ppal.},
	   \ \beta(\varepsilon) \in \mathcal{B}_\alpha \big] \\
0 & \mathrm{otherwise} \\
\end{cases}\ .
\end{equation}

Finally, the outline of the algorithm for calculating the sought derivatives
$\partial \vec{X}^{\,\prime}_\alpha \big(s,f(s)\big) / \partial s^i$
along the constrained subspace $\Sigma$ is:

\begin{enumerate}
\item Calculate the chain $\mathcal{B}_\alpha$ that connects atom
	  $\alpha$ with atom 1 and identify the special cases depicted
	  in fig.~\ref{fig:special_cases}.
\item Calculate the derivatives $\partial f^I / \partial s^i$ by
      solving the system of linear equations in~(\ref{eq:fsolved}).
\item Calculate the geometric derivatives $\partial \vec{X}^{\,\prime}_\alpha
      (r) / \partial s^i$ and $\partial \vec{X}^{\,\prime}_\alpha (r) /
      \partial d^I$, for $I=K+1,\ldots,N$, using eqs.~(\ref{eq:dxp_dr}),
	  (\ref{eq:dxp_dtheta}) and~(\ref{eq:dxp_dphi}).
\item Plug all the calculated quantities into eq.~(\ref{eq:der_xp}) et
      voil\`a.
\end{enumerate}

\section*{Results and Discussion}

In this section, we compare the finite-differences approach (see Methods) to
the new algorithm introduced in this work with two objectives in mind: the
validation of the new scheme, and the identification of the most important
pitfalls of the finite-differences technique, which are absent in the new
method. It is worth stressing again that the method presented here is the
first of its kind, as far as we are aware, and the finite-differences scheme
is just a very natural and straightforward method that is always available
when derivatives need to be calculated. In fact, the pitfalls of finite
differences which we highlight in this section are very well known, although
they have been seldomly presented in the context of molecular force fields.
We hope that this section can be additionally useful to revisit this
classical topic from a new angle.

\begin{figure}[!ht]
\begin{center}
\includegraphics[scale=0.4]{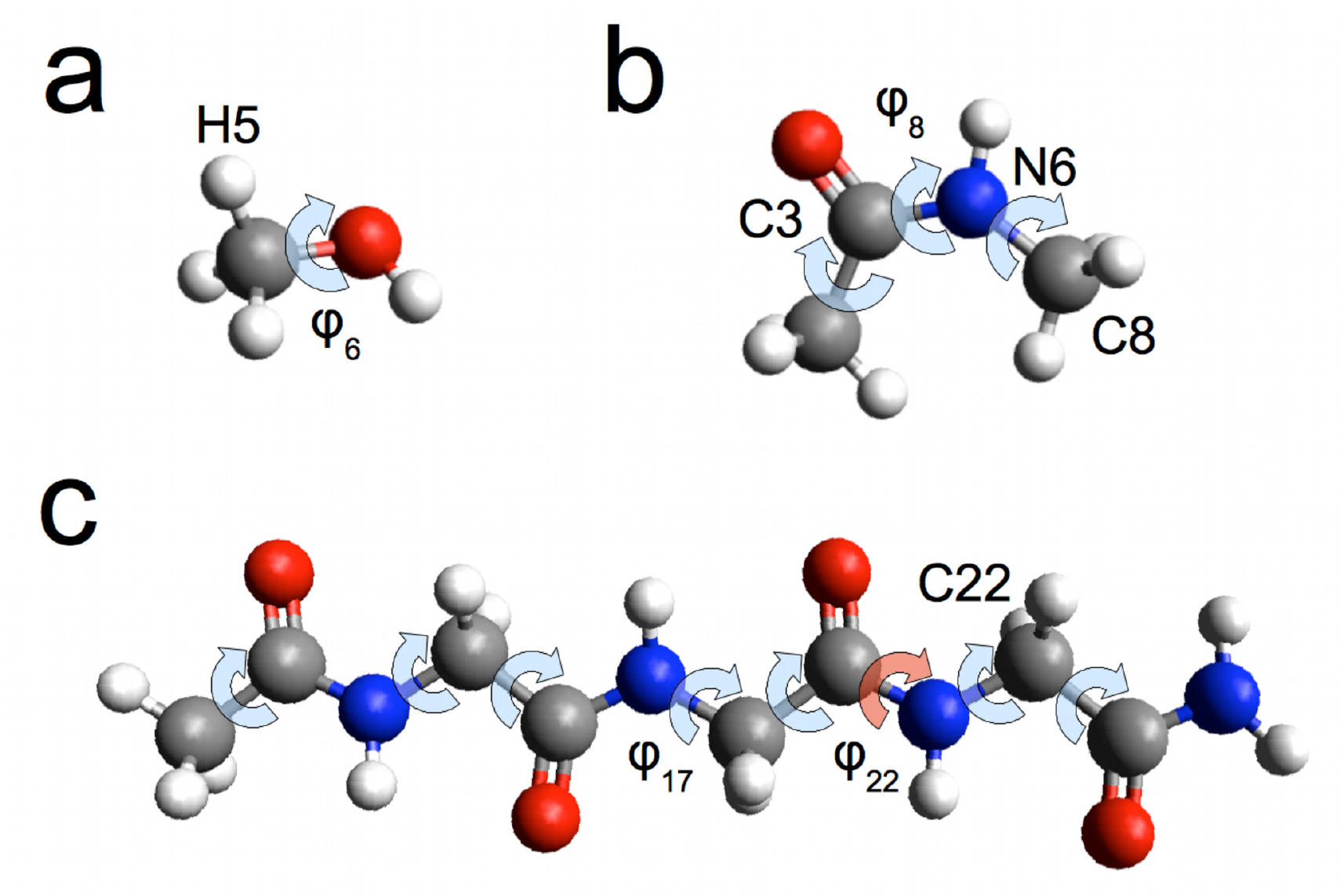}
\end{center}
\caption{
{\bf Molecules used in the numerical calculations in this section.} 
  \textbf{(a)} Methanol, \textbf{(b)} 
  N-methyl-acetamide (abbreviated NMA), and \textbf{(c)}
  the tripeptide N-acetyl-glycyl-glycyl-glycyl-amide (abbreviated GLY3).
  Hydrogens are conventionally white, carbons are grey, nitrogens blue and
  oxygens red. The unconstrained dihedral angles that span the
  corresponding spaces $\mathcal{K}$ are indicated with light-blue arrows,
  and some internal coordinates and some atoms that appear in the discussion 
  are specifically labeled. The constrained dihedral angle $\varphi_{22}$ is
  indicated by a red arrow in GLY3.
}
\label{fig:molecules}
\end{figure}

To these two ends, we have applied the more specific algorithm introduced in
the previous section for the calculation of the derivatives of the Euclidean
coordinates of molecular systems to the three biological species in
fig.~\ref{fig:molecules}: methanol, N-methyl-acetamide (abbreviated NMA), and
the tripeptide N-acetyl-glycyl-glycyl-glycyl-amide (abbreviated GLY3). For
each one of these molecules, a number of dihedral angles describing rotations
around single bonds (and indicated with light-blue arrows in
fig.~\ref{fig:molecules}) have been chosen as the unconstrained internal
coordinates, $s$, spanning the corresponding constrained internal subspace
$\Sigma$. The rest of internal coordinates $d$ (bond lengths, bond angles,
phase dihedrals, and principal dihedrals over non-single bonds) are flexibly
constrained as described in the previous sections. The numeration of the atoms
and the definition of the internal coordinates follow the SASMIC scheme, which
is specially adapted to deal with constrained molecular systems
\cite{Ech2006JCC1}.

For methanol and NMA, due to the small dimensionality of their constrained
subspaces, the working sets of conformations have been generated by
systematically scanning their unconstrained internal coordinates at finite
steps. For methanol, we produced 19 conformations, in which the central
dihedral, $\varphi_6$, ranges from $0^o$ to $180^o$ in steps of $10^o$.
Similarly, the systematic scanning of the unconstrained dihedrals in NMA
produced a set of 588 conformations in which the first and last angles,
$\varphi_6$ and $\varphi_{10}$, range from $0^o$ to $180^o$, and the central
one, $\varphi_8$, ranges from $0^o$ to $330^o$, all in steps of $30^o$. For
GLY3, and in view of the dimensionality of its constrained subspace, 1368
conformations were generated through a Monte Carlo with minimization
procedure.

At each one of these conformations, defined by the value of the unconstrained
internal coordinates $s$, the constrained coordinates $d$ were found by
minimizing the potential energy $V(s,d)$ at fixed $s$, thus enforcing the
constraints $d=f(s)$ described in Methods. Let us remark that this fixing of
the coordinates $s$ is just an algorithmic way of sampling the constrained
subspace defined by the relations $d=f(s)$, and it does not imply that the
coordinates $s$ are constrained; indeed, they could take any value in the set
of conformations, whereas the constrained coordinates $d$ are fixed by the
aforementioned relations. The potential energy and force-field parameters were
taken from the AMBER 96 parameterization \cite{Cor1995JACS,Kol1997Book}, and
local energy minimization with respect to the constrained coordinates was
performed with Gaussian 03 \cite{Gaussian03E01}. At the minimized points, the
Euclidean coordinates, $\vec{Z}^{\,\prime}_\alpha(s) :=
\vec{X}^{\,\prime}_\alpha\big(s,f(s)\big)$, of all atoms in the system-fixed
axes defined in the Methods section were also computed.

In order to find the partial derivatives $\partial \vec{Z}^{\,\prime}_\alpha /
\partial s^i$ at the generated points by finite differences, we produced, for
each conformation in the working sets, $M=K-6$ additional conformations, each
one with a single coordinate $s^i$ displaced to $s^i + \Delta$. After the
re-minimization of the constrained coordinates at the new points, we were in
possession of all the data needed to compute the estimate of the sought
derivative in eq.~(\ref{eq:finite_differences}) for all unconstrained
coordinates. In order to assess the behaviour and accuracy of the
finite-differences approach, we performed these calculations for the values
$\Delta = 0.01^o, 0.05^o, 0.1^o, 0.5^o, 1.0^o,5.0^o,10.0^o$.

On the other hand, to calculate the derivatives $\partial
\vec{Z}^{\,\prime}_\alpha / \partial s^i$ using the new scheme introduced in
Methods, we do not need to perform any additional minimization, but we need to
know the Hessian matrix of the second derivatives of $V(s,d)$ with respect to
the internal coordinates. The Hessian in internal coordinates was calculated
with the Gaussian 03 package \cite{Gaussian03E01}.

\begin{figure}[!ht]
\begin{center}
\includegraphics[scale=0.7]{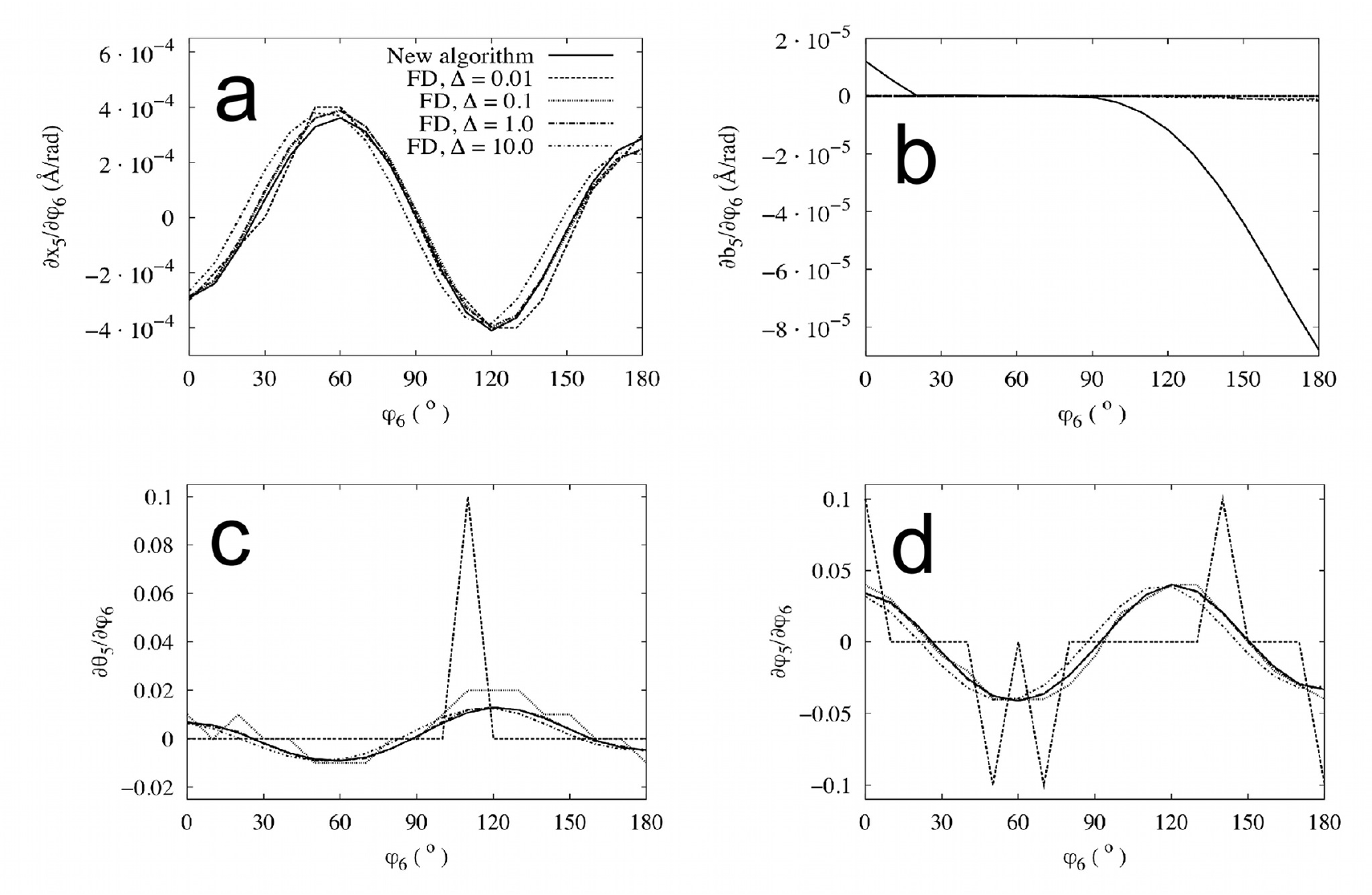}
\end{center}
\caption{
{\bf Derivatives of some selected coordinates of methanol.}
 Derivatives of
 \textbf{(a)} the $x$ coordinate of atom 5 in methanol, \textbf{(b)}
 the bond length $b_5$ associated to it, \textbf{(c)} the bond angle
 $\theta_5$, and \textbf{(d)} the dihedral angle $\varphi_5$ as a function
 of the unconstrained coordinate $\varphi_6$. Both the results of the new
 algorithm and those obtained by finite differences (FD) are depicted.
 The key for the different types of line is the same in the four graphs.
}
\label{fig:methanol_derivatives}
\end{figure}

In order to compare the two methods, we turn first to the smallest system:
methanol. In fig.~\ref{fig:methanol_derivatives}a, we can see the value of the
derivative $\partial x_5 / \partial \varphi_6$ of the $x$-coordinate (in the
`primed' axes, but we drop the prime from now on) of hydrogen number 5 (see
fig.~\ref{fig:molecules}) with respect to the unconstrained dihedral angle
$\varphi_6$ that describes the rotation of the alcohol group with respect to
the methyl one. We can see that the agreement between the new algorithm and
the finite-differences approach is good but not perfect, and that the
discrepancy between the two is larger for the smallest ($0.01^o$) and largest
($10.0^o$) values of $\Delta$ depicted in the graph.

To track the source of this difference, we can take a look at
eq.~(\ref{eq:derZ1bis}), which gives the derivative $\partial
\vec{Z}^{\,\prime}_\alpha / \partial s^i$ as a function of simple,
`geometrical' terms, $\partial \vec{X}^{\,\prime}_\alpha / \partial s^i$ and
$\partial \vec{X}^{\,\prime}_\alpha / \partial d^I$, and the numerical
derivatives $\partial f^I / \partial s^i$. Of course, the choice of one method
or another does not affect the former, but only the latter. In the particular
case of $\partial x_5 / \partial \varphi_6$ in
fig.~\ref{fig:methanol_derivatives}a, if we remove the terms that are zero
according to the rules in eqs.~(\ref{eq:dxp_dr}), (\ref{eq:dxp_dtheta})
and~(\ref{eq:dxp_dphi}), eq.~(\ref{eq:derZ1bis}) becomes
\begin{eqnarray}
\label{eq:dx5dphi6_methanol}
\frac{\partial x_5}{\partial \varphi_6} & = &
  \frac{\partial x_5}{\partial b_2} \frac{\partial b_2}{\partial \varphi_6}
+ \frac{\partial x_5}{\partial b_3} \frac{\partial b_3}{\partial \varphi_6}
+ \frac{\partial x_5}{\partial \theta_3}
  \frac{\partial \theta_3}{\partial \varphi_6}
  \nonumber \\
 & & \mbox{} + \frac{\partial x_5}{\partial b_5} 
  \frac{\partial b_5}{\partial \varphi_6}
+ \frac{\partial x_5}{\partial \theta_5}
  \frac{\partial \theta_5}{\partial \varphi_6}
+ \frac{\partial x_5}{\partial \varphi_5}
  \frac{\partial \varphi_5}{\partial \varphi_6} \ .
\end{eqnarray}

The numerical derivatives appearing in this expression that are related to the
three constrained coordinates associated to atom 5 are shown in
figs.~\ref{fig:methanol_derivatives}b, \ref{fig:methanol_derivatives}c and
\ref{fig:methanol_derivatives}d, respectively, where we can see that the
discrepancy between the new algorithm and the finite-differences approach is
more significant. For the bond angle $b_5$ in
fig.~\ref{fig:methanol_derivatives}b, we see that the derivative predicted by
finite differences is close to zero for all values of $\varphi_6$ and for all
the tested $\Delta$s, while the behaviour given by the new algorithm is more
rich and substantially different. This large discrepancy is produced by the
fact that bond lengths are very stiff coordinates in the energy function that
we have used here, together with the default precision of the floating point
numbers provided by Gaussian 03 outputs. In table~\ref{tab:methanol_hard}, we
can see indeed that the last significant figure of bond length $b_5$ only
starts to change for $\Delta=5.0^o$, which makes any algorithm based on finite
differences very unreliable for this particular quantity if small values of
$\Delta$ are used. The bond angles and dihedral angles, on the other hand, are
somewhat less stiff than bond lengths, as it can also be seen in
tab.~\ref{tab:methanol_hard}. This makes their derivatives by finite
differences more reliable, as one can observe in
figs.~\ref{fig:methanol_derivatives}c and \ref{fig:methanol_derivatives}d,
where the discrepancy with the new method is apparent for small $\Delta$, but
becomes gradually smaller as we increase it. Of course, since, in the new
method presented in this work, all quantities are computed at the
non-displaced point $\Delta=0^o$, the problem regarding the number of
significant figures does not appear. It is also worth remarking that, in the
case of finite differences, the point in which this issue will appear depends
on the number of bits used to represent coordinates, but it will always appear
for some small enough value of $\Delta$.

\begin{figure}[!ht]
\begin{center}
\includegraphics[scale=0.7]{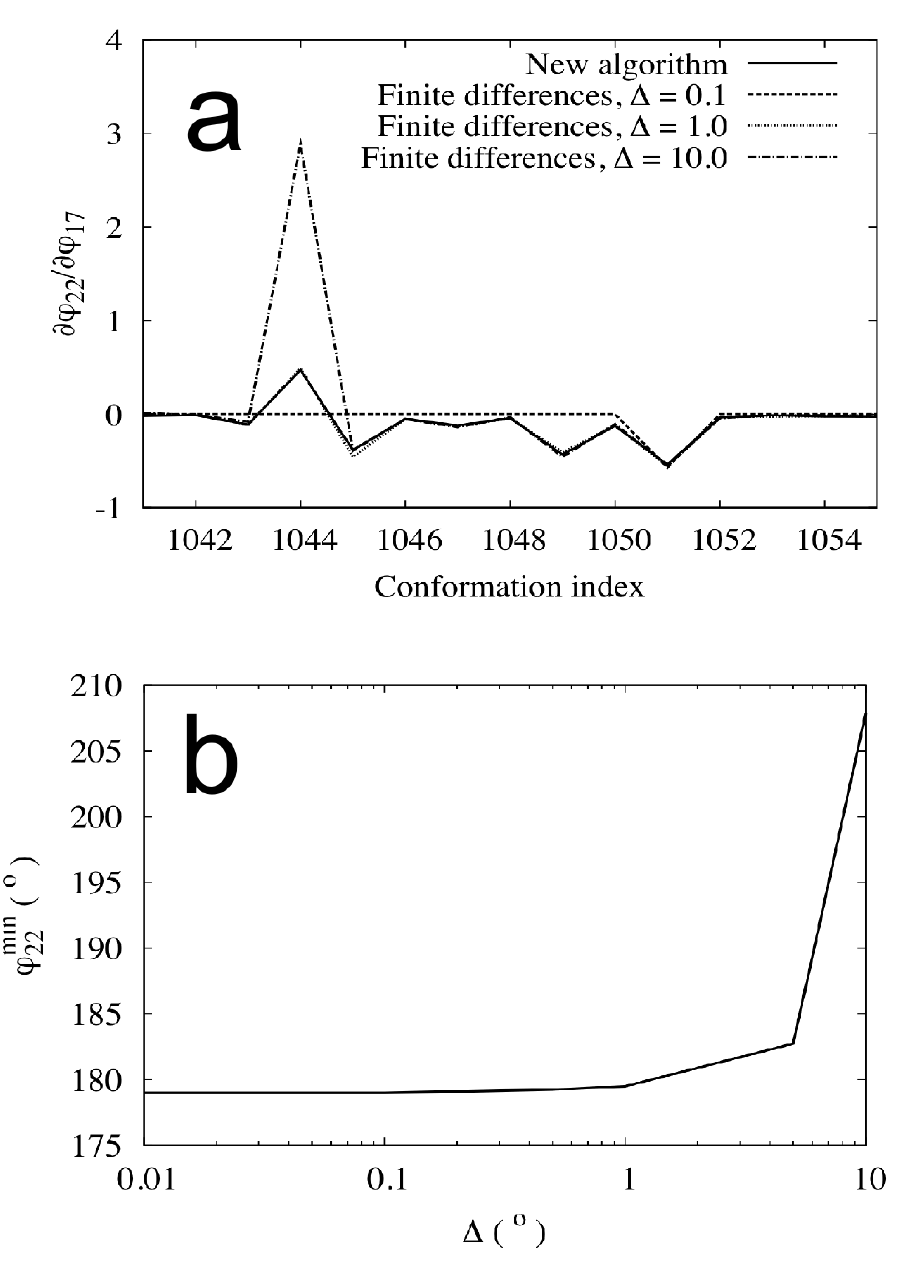}
\end{center}
\caption{
{\bf Metastability of the local minima in GLY4.} \textbf{(a)} Derivative
 $\partial \varphi_{22} / \partial \varphi_{17}$ of the constrained dihedral
 angle $\varphi_{22}$, describing a peptide bond rotation in GLY3, with
 respect to the unconstrained coordinate $\varphi_{17}$ for a selected set of
 conformations in the working set. \textbf{(b)} Minimum-energy value
 of the constrained dihedral angle $\varphi_{22}$ in the conformation 1044
 of GLY3 for different values of the displacement $\Delta$ in the
 unconstrained coordinate $\varphi_{17}$.
}
\label{fig:metastable}
\end{figure}

As we noticed in fig.~\ref{fig:methanol_derivatives}a, also in the case of the
constrained internal coordinates the difference between the two methods starts
to grow again when $\Delta$ reaches $5.0^o$ or $10.0^o$. This is easily
understood if we think that only in the $\Delta \to 0$ limit the estimate in
eq.~(\ref{eq:finite_differences}) converges to the actual value of the partial
derivative. In fact, as the complexity of the system increases, the error
introduced at large $\Delta$ may come not only from continuous changes in the
location of the constrained minima, but also, as fig.~\ref{fig:metastable}
suggests, it may occur that, at a certain value of $\Delta$, the very
\emph{identity} of the minima is altered, thus introducing potentially larger
errors. In fig.~\ref{fig:metastable}a, we can see that the derivative
$\partial \varphi_{22} / \partial \varphi_{17}$ in GLY3 presents an unusually
large error at the conformation 1044. In fig.~\ref{fig:metastable}b, we see
that the minimum-energy value of $\varphi_{22}$, which is the dihedral angle
associated to carbon 22, describing the rotation around a given peptide bond
(see fig.~\ref{fig:molecules}c), presents an abrupt change when $\Delta$
reaches $10^o$. If we think that the energy landscape of GLY3 is indeed a
complex and multidimensional one, it is not difficult to imagine that, as we
change $\varphi_{17}$, i.e., as we increase $\Delta$, the energy landscape is
so altered that some minima disappear, some other appear, and the energy
ordering among them is changed. In such a case, the structures found by the
minimization procedure will be rather different between, say, $\Delta=0^o$ and
$\Delta=10^o$, thus producing a large error in the derivatives calculated by
finite differences. Again, the new algorithm, which only uses quantities
calculated at $\Delta=0^o$, does not suffer from this drawback.

\begin{figure}[!ht]
\begin{center}
\includegraphics[scale=0.7]{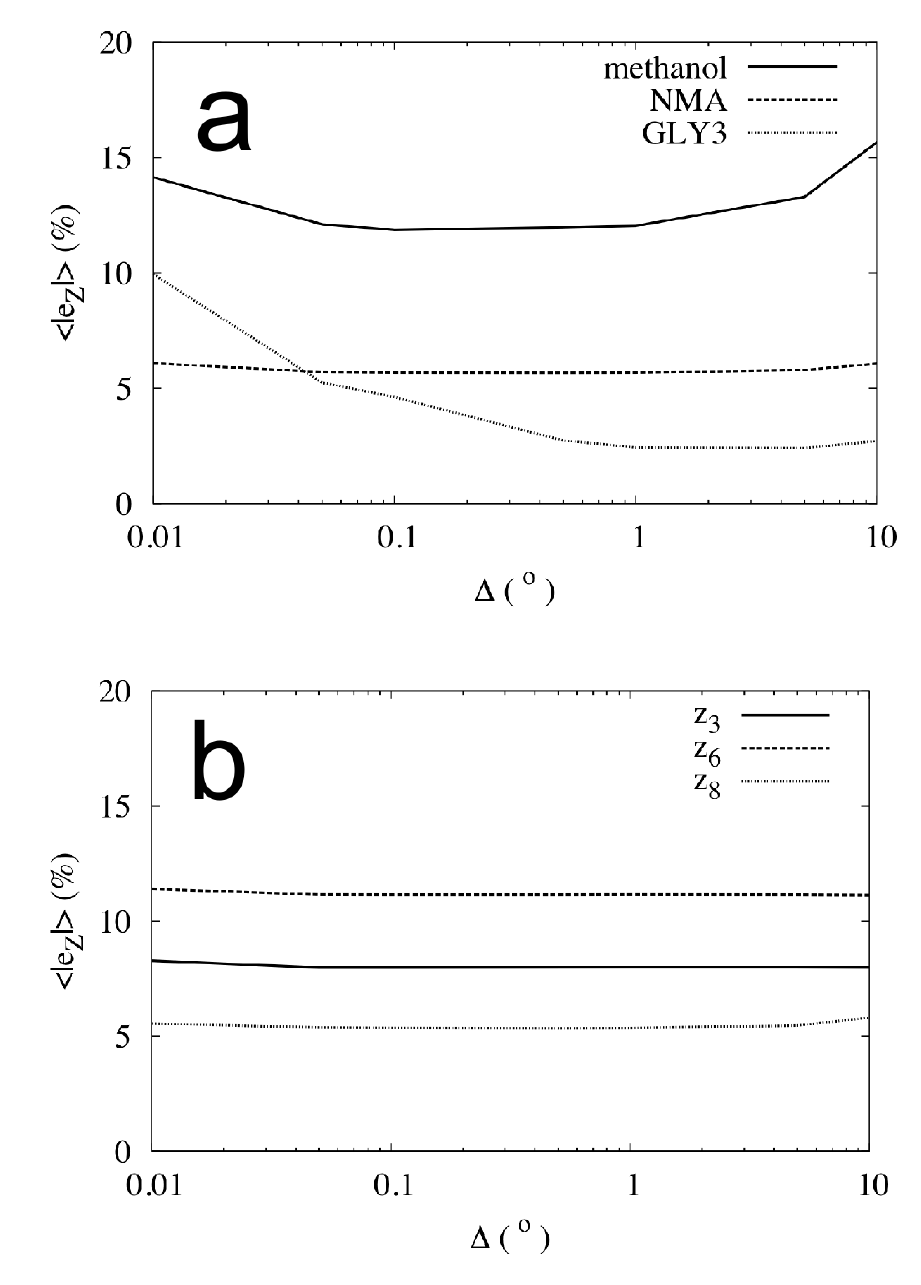}
\end{center}
\caption{
{\bf Dependence of the error as a function of $\Delta$.} Average normalized 
  error
  in the derivatives by finite differences as a function of $\Delta$ (see
  the text for a more precise definition). \textbf{(a)} Error averaged to
  all conformations and all atoms of the three molecular systems studied.
  \textbf{(b)} Error averaged to all conformations of the $z$-coordinate
  of three particular $1^\mathrm{st}$-row atoms in NMA.
}
\label{fig:delta_system_tuning}
\end{figure}

To sum up, the finite-differences method contains two sources of error which
the new method does not present: one at small values of $\Delta$, related to
the finite precision of the floating point numbers representing the internal
coordinates, and the other at larger values of $\Delta$, stemming from the
very definition of the partial derivative by finite differences, and
aggravated by the complexity of the energy landscapes of large systems. If the
derivatives are to be calculated using finite differences, an optimal value of
$\Delta$ must be chosen in each case so that the possible error is minimized.
However, already in the simple example of methanol, we saw that the
derivatives of different observables, in the same system, may behave
differently as we change $\Delta$ (compare the bond length derivative in
fig.~\ref{fig:methanol_derivatives}b with that of the angles in
figs.~\ref{fig:methanol_derivatives}c and~\ref{fig:methanol_derivatives}d). In
fig.~\ref{fig:delta_system_tuning}, we additionally see that the search for
the optimal $\Delta$ may be further complicated by the fact that the behaviour
found also depends (strongly) on the system studied, and, in the case of the
derivatives of the Euclidean coordinates, on the position of the atom in the
molecule.

In fig.~\ref{fig:delta_system_tuning}a, we have plotted the normalized average
of the absolute value of the error in the derivatives of the Euclidean
coordinates, $\langle |e_Z| \rangle$, as a function of $\Delta$ for the
three molecular systems studied. This quantity is defined, for a given
unconstrained coordinate $s^i$, as
\begin{eqnarray}
\label{eq:eZ}
\langle |e_Z| \rangle(\Delta) & := & \frac{100}{N_\mathrm{c}(3n-6)} \\
  & \times & \sum_{m=1}^{N_\mathrm{c}} \sum_{\mu=1}^{3n}
  \frac{1}{\delta^\mu_i} \left| 
  \left( \frac{\partial Z^\mu}{\partial s^i} \right)^m_\mathrm{FD}
  \!\!\!(\Delta) -
  \left( \frac{\partial Z^\mu}{\partial s^i} \right)^m_\mathrm{NA}
  \right| \ , \nonumber 
\end{eqnarray}
where the index $m$ indicates the conformation in the working set, running
from 1 to $N_\mathrm{c}$, FD stands for `finite differences', NA for
`new algorithm', and $\delta^\mu_i$ is a normalizing quantity for each
coordinate $x^\mu$ chosen as
\begin{equation}
\label{eq:deltamui}
\delta^\mu_i :=
 \max_m \left( \frac{\partial Z^\mu}{\partial s^i} \right)^m_\mathrm{NA} -
 \min_m \left( \frac{\partial Z^\mu}{\partial s^i} \right)^m_\mathrm{NA} \ .
\end{equation}

The graphics in fig.~\ref{fig:delta_system_tuning}a of this quantity
correspond to the unconstrained dihedral angles $\varphi_6$, $\varphi_8$ and
$\varphi_{17}$ of methanol, NMA and GLY3, respectively (see
fig.~\ref{fig:molecules}). We observe that the average error as a function of
$\Delta$ presents significantly different behaviours in the three molecules,
never being smaller than a 2\%. Additionally, in
fig.~\ref{fig:delta_system_tuning}b, we show the same error but this time
individualized to the $z$-coordinate of three different $1^\mathrm{st}$-row
atoms of NMA: C3, N6 and C8. Although the overall behaviour of the error is
similar for the three atoms, its size is not.

All in all, we see that the need to tune for the optimal $\Delta$ in the
finite-differences approach not only produces unavoidable errors, but also it
must be done in a per-system, per-observable basis, clearly complicating and
limiting the use of this technique. The new algorithm, on the other hand, is
only affected by the source of error related to the accuracy with which the
Hessian matrix of the potential energy can be calculated and inverted; apart
from this, which is a general drawback of any method implemented in a
computer, its mathematical definition is `exact', in the sense that it does
not contain any tunable parameter, like $\Delta$, that must be adjusted for
optimal accuracy in each particular problem.

Also, and more importantly (since the failure of finite differences was indeed
predictable) the good coincidence between the newly introduced, somewhat more
involved method and the straightforward finite-differences scheme for the
smallest system and in some intermediate range of values of $\Delta$ allows us
to regard the new scheme as validated and error-free.

Finally, despite what we discussed in the Methods section, namely, that we
have not pursued here the numerical optimization of the algorithm introduced,
being our main interest to present the general theoretical concepts and to
show that the new method is exact and reliable, we close this section with an
example of a toy system to provide a clue that the new technique is at least
feasible. Before introducing the toy system it is worth noting that the
examples tackled in this section are just particular cases, but the technique
can be used in different systems and with different potential energy
functions. When looking at the computer costs presented below, the reader
should bear in mind that they may be not very significant (due to the
aforementioned lack of optimization) and not very relevant (due to the choice
of a small toy system and a given potential energy function). Of course, if
any production runs using the new algorithm are attempted, a thorough
numerical optimization and assessment should be performed, which we deem to be
a very important next step of our work.

\begin{figure}[!ht]
\begin{center}
\includegraphics[scale=0.2]{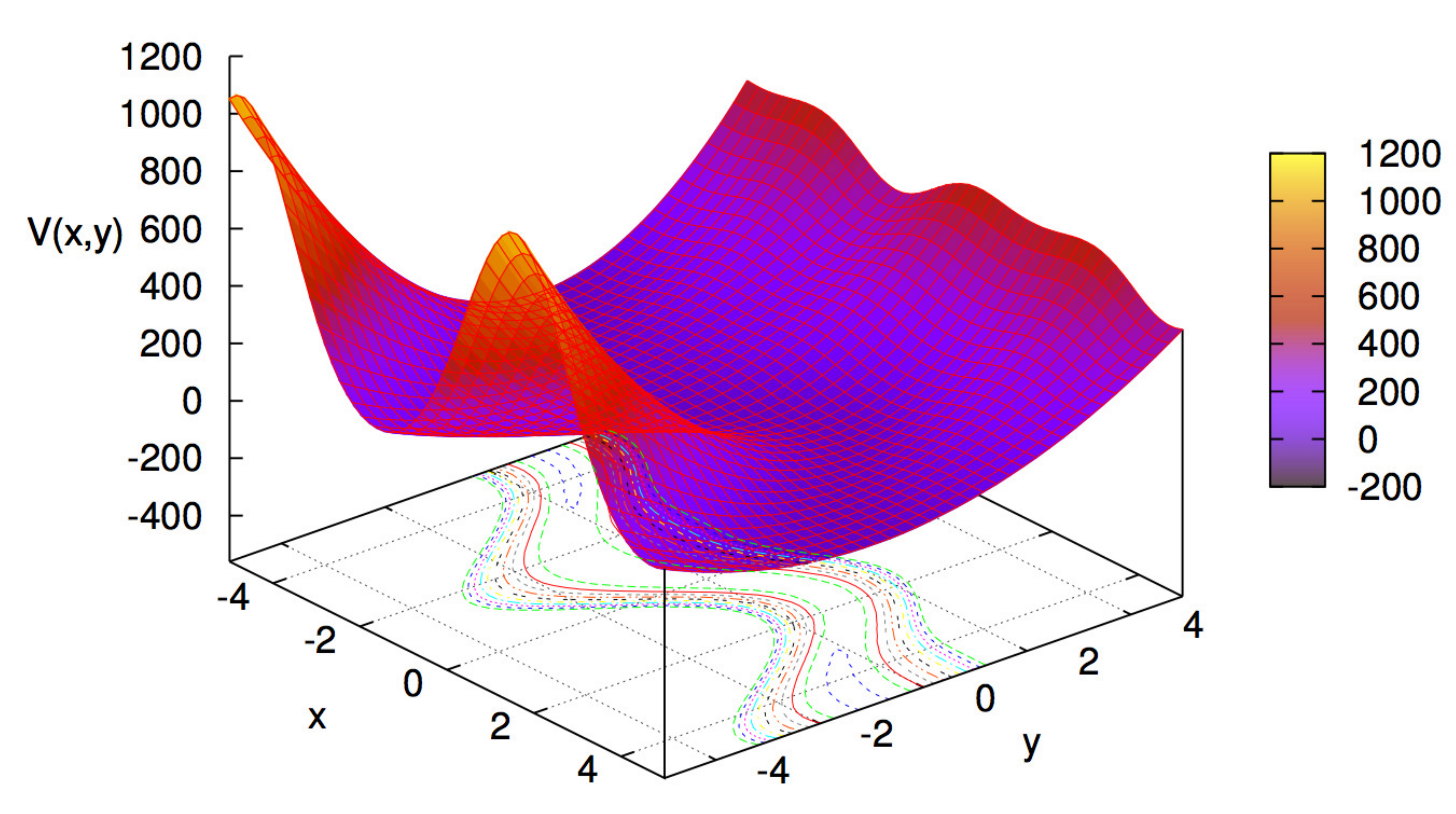}
\end{center}
\caption{
  {\bf Potential energy of the toy system in 
  eq.~(\ref{eq:V_toy}).} The range of $x$ and $y$ corresponds to the one
  explored in this work. Contour level lines and colour level indication
  in the surface have been added for visual comfort. All units are
  arbitrary.
}
\label{fig:V_toy}
\end{figure}

The toy system is a 2-dimensional one, with positions $x$ and
$y$, and the following potential energy (see fig.~\ref{fig:V_toy}):
\begin{equation}
\label{eq:V_toy}
V(x,y)=\frac{1}{2}K(2+\sin x)(y - \sin x)^2 + \tanh (xy) \ .
\end{equation}

If we take a large enough $K$, say $K=20$, we see that the
system will present a strong oscillatory motion in the $y$ coordinate, around
approximately $y=\sin x$ (but not exactly, since the term $\tanh (xy)$
slightly modifies the position of the minimum), and with harmonic constant
approximately equal to $K(2+\sin x)$. In the spirit of this work, since, due
to energetic reasons, the value of $y$ will seldomly move far away from the
value that minimizes $V(x,y)$ for each $x$, denoted by $f(x)$ and implicitly
defined by the following equation:
\begin{equation}
\label{eq:der_V_toy}
\frac{\partial V}{\partial y}\big(x,f(x)\big) =
 K (2+\sin x) \big(f(x)-\sin x\big) + \frac{x}{\cosh^2 \big(xf(x)\big)}
 = 0 \ ,
\end{equation}
we can kill this oscillatory motion by assuming that a flexible constraint
$y=f(x)$ exists. In such a case, $x$ plays the role of the whole set of
unconstrained coordinates $s$ in the general formalism, and $y$ plays the role
of the whole set of constrained coordinates $d$.

Now, if we perform a `molecular dynamics' of this system, then
we may need at some point to compute the derivative with respect to $x$ of
some observable $X(x,y)$ restricted to the constrained subspace $Z(x) :=
X\big(x,f(x)\big)$ (for example, we may need this to calculate mass-metric
tensor corrections at each time step \cite{Ech2006JCC2}). We can do so by
using finite differences or the new technique introduced in this work. As we
discussed in the Methods section, for both approaches we will need to perform
a minimization of $V(x,y)$ at each fixed $x$ in order to find $f(x)$ and $Z(x)
:= X\big(x,f(x)\big)$, hence, being this step common, we will not consider it
for the assessment of the differences in computational cost between the two
methods. The additional computations that will decide which method is faster
are:

\begin{itemize}

\item \textbf{For finite differences:} Choose a displaced
value of the unconstrained coordinate $x + \Delta x$, minimize $V(x + \Delta
x, y)$ with respect to $y$ in order to find $f(x + \Delta x)$, as well as
$X\big(x + \Delta x, f(x + \Delta x)\big)$, and finally calculate the
finite-differences estimation of the sought derivative:
\begin{equation}
\label{eq:FD_V_toy}
\frac{\partial Z}{\partial x}(x) \simeq
\frac{X\big(x + \Delta x, f(x + \Delta x)\big)-X\big(x,f(x)\big)}
     {\Delta x} \ .
\end{equation}

\item \textbf{For the new method:} Calculate the objects in
eq.~(\ref{eq:fsolved}), perform the required inversion to find $\partial
f/\partial x$, calculate the objects in eq.~(\ref{eq:derZ1}), and finally find
$\partial Z/\partial x$ using this last expression. In this simple case, all
the objects to be computed are:
\begin{equation}
\label{eq:V_toy_objects}
\frac{\partial^2 V}{\partial y^2}\big(x,f(x)\big) \ , \quad
\frac{\partial^2 V}{\partial x \partial y}\big(x,f(x)\big) \ , \quad
\frac{\partial X}{\partial x}\big(x,f(x)\big) \ , \ \mathrm{and} \quad
\frac{\partial X}{\partial y}\big(x,f(x)\big) \ .
\end{equation}

\end{itemize}

The second-order derivatives of the potential energy can be
easily calculated:
\begin{subequations}
\label{eq:der2_V_toy}
\begin{align}
\frac{\partial^2 V}{\partial y^2}(x,y) & =
 K(2 + \sin x) - \frac{2x^2 \tanh(xy)}{\cosh^2(xy)} \ ,
 \label{eq:der2_V_toy_a} \\
\frac{\partial^2 V}{\partial x \partial y} & =
 K \cos x \big(y - 2 [1 + \sin x]\big) + \frac{1 - 2xy \tanh(xy)}{\cosh^2(xy)}
 \ , \label{eq:der2_V_toy_b}
\end{align}
\end{subequations}
and we can use them to find the derivative of $f(x)$ through 
eq.~(\ref{eq:fsolved}):
\begin{equation}
\label{eq:derf_V_toy}
\frac{\partial f}{\partial x}(x) := -
 \left[\frac{\partial^2 V}{\partial x \partial y}\big(x,f(x)\big)\right] \Big/
 \left[\frac{\partial^2 V}{\partial y^2}\big(x,f(x)\big)\right] \ .
\end{equation}

The particularization of eq.~(\ref{eq:derZ1}) to this simple
case is
\begin{equation}
\label{eq:derZ1_V_toy}
\frac{\partial Z}{\partial x}(x) =
 \frac{\partial X}{\partial x}\big(x,f(x)\big) +
 \frac{\partial X}{\partial y}\big(x,f(x)\big)
 \frac{\partial f}{\partial x}(x) \ .
\end{equation}

In this section, for illustrative purposes, we have chosen
a simple observable $X(x,y)$:
\begin{equation}
\label{eq:X_V_toy}
X(x,y):=\sqrt{x^2+y^2} \ ,
\end{equation}
i.e., the distance of the particle to the origin of coordinates. Hence,
the remaining objects that we need to compute in order to apply the new
technique to this problem are
\begin{subequations}
\label{eq:derX_V_toy}
\begin{align}
\frac{\partial X}{\partial x}(x,y) & = \frac{x}{\sqrt{x^2+y^2}} \ ,
 \label{eq:derX_V_toy_a} \\
\frac{\partial X}{\partial y}(x,y) & = \frac{y}{\sqrt{x^2+y^2}} \ .
 \label{eq:derX_V_toy_b}
\end{align}
\end{subequations}

We have calculated $\partial Z / \partial x$ using both
techniques for 11 different values of $x=-5.0,-4.0,\ldots,4.0,5.0$. This
calculation has been performed in a desktop iMac with a 2.66 GHz Intel Core 2
Duo processor and 4GB of 1067 MHz DD3 RAM memory, running MacOSX Snow Leopard.
The same compilation-time optimizations have been used in the two cases, and
the common times have been subtracted as indicated before. It is also worth
remarking that we have used Brent's method \cite{Pre2007Book} for minimizing
$V(x,y)$, and a different choice will change the comparison. In these
conditions, the new technique has proved approximately one order of magnitude
faster than finite differences, elapsing $0.13$ $\mu$s/point vs. $1.08$
$\mu$s/point.

In summary, in this work, we have introduced a new, exact, parameter-free
method for computing the derivatives of physical observables in systems with
flexible constraints. The new algorithm has been numerically validated in
small molecules against its most natural alternative, finite differences. In
doing so, numerous pitfalls of the latter method have been demonstrated, all
arising from the fact that it contains a tunable parameter that has to be
optimally adjusted in each particular problem at hand. In a number of
numerical experiments, we have shown that the finite-differences approach
contains two unavoidable sources of error that are not present in the new
method: On the one hand, the finite number of significant figures used to
represent, in computers, the values of the optimized coordinates, together
with the fact that these constrained coordinates are typically very stiff,
make the changes in this quantities often unobservable or at least badly
resolved, thus rendering the finite-differences derivatives unreliable for
small values of the displacement parameter $\Delta$. On the other hand, the
very fact that finite-differences derivatives only converge to the true ones
for $\Delta \to 0$, complicated with the possibility that the energy
landscapes of complex molecular systems may significantly change their
structure when the unconstrained coordinates are displaced, introduce new
errors as $\Delta$ increases. These two sources of errors combined make
compulsory the search of an optimal value of $\Delta$ in each particular case,
and also establish a minimum error below which is not possible to go, as it
can be seen in fig.~\ref{fig:delta_system_tuning}. Also, using a simple toy
system, we have shown that the new technique can be faster than finite
differences in certain situations. The new method introduced here, and it is
already being successfully used in a number of works in progress in our group
to compute the correcting terms appearing in the equilibrium probability
distribution when flexible constraints are imposed on the system
\cite{Ech2011Sub2}. Moreover, given the almost ubiquitous occurrence of the
concept of constraints all throughout the fields of computational physics and
chemistry, it is expected that the method described in this work will find
many applications in present and future problems. Some examples have been
already mentioned in the introduction, notably the case of ground-state
Born-Oppenheimer MD \cite{Alo2008PRL,And2009JCTC} (using, e.g., Hartree-Fock
\cite{Ech2007MP}), which can be regarded as a flexibly constrained problem in
which the soft coordinates are the nuclear positions $R$, the hard ones are
the electronic orbitals $\psi$, and the function to be minimized is the
expected value $\langle \Psi | \hat{H}_e(R) | \Psi \rangle$ of the
$R$-dependent electronic Hamiltonian in the $N$-electron Slater determinant
$\Psi$.

\section*{Acknowledgments}

We thank the reviewers of the manuscript for their insightful comments that
have contributed to improve the final version. We aknowledge funding from the
grants FIS2009-13364-C02-01 (MICINN, Spain), Grupo de Excelencia
``Biocomputaci\'on y F\'{\i}sica de Sistemas Complejos'', E24/3 (Arag\'on
region Government, Spain), ARAID and Ibercaja grant for young researchers
(Spain). P. G.-R. has been supported by a JAE-Predoc scholarship (CSIC,
Spain).

\section*{Figure Legends}

\section*{Tables}

\begin{table}[!ht]
\caption{
\bf{Stiffness of the constrained coordinates in methanol}}
\begin{tabular}{r@{.}l|ccc}
\multicolumn{2}{l}{$\Delta$ ($^o$)} &
             $b_5$ (\AA)  & $\theta_5$ ($^o$) & $\varphi_5$ ($^o$) \\
\hline
	0&0        & 1.090694 & 109.403           & 119.296            \\
	0&01       & 1.090694 & 109.404           & 119.296            \\
	0&05       & 1.090694 & 109.404           & 119.297            \\
	0&1        & 1.090694 & 109.405           & 119.299            \\
	0&5        & 1.090694 & 109.409           & 119.312            \\
	1&0        & 1.090694 & 109.415           & 119.329            \\
	5&0        & 1.090693 & 109.462           & 119.474            \\
	10&0       & 1.090692 & 109.525           & 119.671
\end{tabular}
\begin{flushleft} Values of the constrained coordinates
 associated to atom 5 of methanol for different displacements $\Delta$ in
 the unconstrained coordinate $\varphi_6$. The values correspond to the
 conformation with $\varphi_6=110^o$, and the number of significant
 figures presented is the default one provided by Gaussian 03.
\end{flushleft}
\label{tab:methanol_hard}
\end{table}

\end{document}